\documentclass[pmlr]{jmlr}% new name PMLR (Proceedings of Machine Learning)

\usepackage{longtable}% for long tables

 \usepackage{booktabs}

\usepackage[load-configurations=version-1]{siunitx} % newer version

\makeatletter
\def\set@curr@file#1{\def\@curr@file{#1}} %temp workaround for 2019 latex release
\makeatother

\theorembodyfont{\upshape}
\theoremheaderfont{\scshape}
\theorempostheader{:}
\theoremsep{\newline}

\jmlrvolume{182}
\jmlryear{2022}
\jmlrworkshop{Machine Learning for Healthcare}

 \makeatletter
\def\@fnsymbol#1{\ensuremath{\ifcase#1\or \dagger\or \ddagger\or
   \mathsection\or \mathparagraph\or \|\or **\or \dagger\dagger
   \or \ddagger\ddagger \else\@ctrerr\fi}}
    \makeatother

\title[Multi-modal fusion with clinical time-series data and chest X-ray images]{MedFuse: Multi-modal fusion with clinical time-series data and chest X-ray images}

\author{\Name{Nasir Hayat}\thanks{Currently at G42 Healthcare.}
      \Email{nasirhayat6160@gmail.com}\\ 
      \addr Engineering Division\\
      NYU Abu Dhabi\\
      Abu Dhabi, UAE 
      \AND
      \Name{Krzysztof J. Geras}
      \Email{k.j.geras@nyu.edu }\\ 
      \addr Department of Radiology\\
      NYU Grossman School of Medicine\\
      New York, NY, USA
      \AND
      \Name{Farah E. Shamout}
      \Email{farah.shamout@nyu.edu}\\ 
      \addr Engineering Division\\
      NYU Abu Dhabi\\
      Abu Dhabi, UAE} 

\begin{document}

\maketitle

\begin{abstract}
 Multi-modal fusion approaches aim to integrate information from different data sources. Unlike natural datasets, such as in audio-visual applications, where samples consist of ``paired'' modalities, data in healthcare is often collected asynchronously. Hence, requiring the presence of all modalities for a given sample is not realistic for clinical tasks and significantly limits the size of the dataset during training. In this paper, we propose \texttt{MedFuse}, a conceptually simple yet promising LSTM-based fusion module that can accommodate uni-modal as well as multi-modal input. We evaluate the fusion method and introduce new benchmark results for in-hospital mortality prediction and phenotype classification, using clinical time-series data in the MIMIC-IV dataset and corresponding chest X-ray images in MIMIC-CXR. Compared to more complex multi-modal fusion strategies, \texttt{MedFuse} provides a performance improvement by a large margin on the fully paired test set. It also remains robust across the partially paired test set containing samples with missing chest X-ray images. We release our code for reproducibility and to enable the evaluation of competing models in the future. 
\end{abstract}
% aggregates encoded features of partially paired datasets. 
% our work emphasizes that due consideration should be given to fusing partially paired medical data modalities for downstream prediction tasks.

\section{Introduction}

Humans perceive the world through multi-modal data~\citep{ngiam2011multimodal}. To date, most of the successful models learning from perceptual data in healthcare are uni-modal, i.e. they rely on a single data modality~\citep{Huang2020_survey}. Multi-modal learning has been widely explored in the context of audio-visual applications~\citep{vaezi20mmtm} and natural image datasets~\citep{zellers2021merlot, zsd}, but less so in healthcare. The main goal of multi-modal fusion is to exploit relevant information from different modalities to improve performance in downstream tasks~\citep{baltruvsaitis2018multimodal}. Multi-modal fusion strategies can be characterized as early, joint, or late fusion~\citep{Huang2020_survey}. The joint fusion paradigm is the most promising, since its core idea is to model interactions between the representations of the input modalities.

We highlight two main challenges facing multi-modal joint fusion in healthcare. First, many of the state-of-the-art approaches make a strong assumption that all modalities are available for every sample during training, inference, or both~\citep{daft}. Although some clinical studies follow suit of this assumption~\citep{Huang2020_survey}, obtaining paired data is not feasible since daily clinical practice produces heterogeneous data with varying sparsity. For example, physiological data is more frequently collected than chest X-ray images in the Intensive Care Unit (ICU) setting. These two modalities are the key focus of our study because they play a very important role in clinical prediction tasks~\citep{benchhmark, Lohan2019}. Developing a unified fusion model for those two modalities also presents its own challenges as they (i) have significantly different input dimensions, (ii) require modality-specific feature extractors due to differences in information and noise content~\citep{nagrani2021attention}, and (iii) are not temporally aligned and hence cannot be paired easily. Considering those challenges, our primary aim is to propose a fusion architecture that can deal with partially paired data, in order to achieve favorable performance in downstream prediction tasks.

%Since medical data is sparse, our primary aim is to deal with missing data through a flexible fusion architecture aligned with clinical practice, . 

\begin{figure}[h!]
    \centering
    \includegraphics[width=0.7\textwidth]{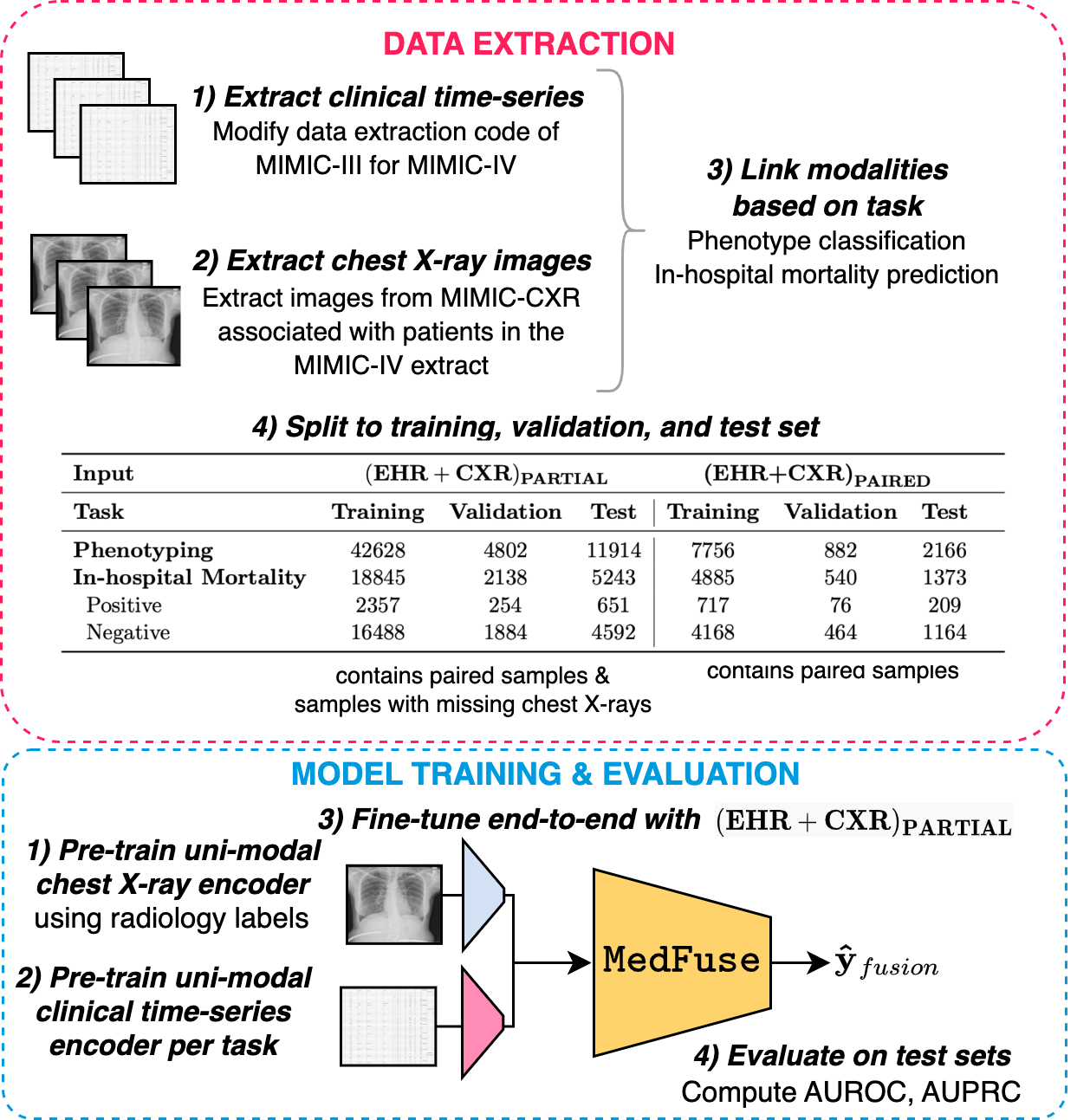} \vspace{-3mm}
    \caption{\small\textbf{Overview of the proposed work.} We first extract and link the datasets from MIMIC-IV and MIMIC-CXR based on the task definition (i.e., inhospital mortality prediction, or phenotype classification). The data splits of the training, validation, and test sets are summarized for each task, and the prevalence of positive and negative labels for in-hospital mortality is shown. Phenotype classification involves 25 labels as shown in Table~\ref{tab:phenotype_wise}.} \vspace{-9mm}
    \label{fig:overview-of-work}
\end{figure}

The second challenge is that there are no well-studied publicly available multi-modal clinical benchmarks. Therefore, most studies rely on a single data modality to perform clinical prediction tasks~\citep{benchhmark}, or use privately curated multi-modal datasets~\citep{Huang2020_survey}. Here, our secondary aim is to introduce new multi-modal benchmark results for two popular clinical prediction tasks using the publicly available Medical Information Mart for Intensive Care (MIMIC)-IV~\citep{mimic4} and MIMIC-CXR~\citep{mimiccxrjpg} datasets, and we also release the code for reproducibility. We compare our approach to vanilla early and joint fusion as well as open-source state-of-the-art joint fusion approaches~\citep{vaezi20mmtm,daft}. In summary, we make the following contributions:
% [itemsep=-4pt, topsep=-3pt]
\begin{itemize}
\item We propose \texttt{MedFuse}, a new LSTM-based~\citep{hochreiter1997long} multi-modal fusion approach. Conventional joint fusion strategies concatenate feature representations of multiple modalities as a single feature representation, and then process that concatenated representation for downstream tasks, such as using a classifier. On the contrary, we treat the multi-modal representation as a sequence of uni-modal representations (or tokens), such that the fusion module aggregates these representations through the recurrence mechanism of LSTM. We assume a sequential structure to leverage the recurrent inductive bias of LSTM and to handle input sequences of variable length, in case of a missing modality. The fusion module is agnostic to the architecture of the modality-specific extractors and can handle missing data during training and inference.
% we concatenate the feature vectors as a sequence of vectors that are dynamically accumulated by the fusion module.
\item To evaluate the proposed approach, we link two open-access real-world datasets: MIMIC-IV~\citep{mimic4}, which contains clinical time-series data collected in the ICU, and MIMIC-CXR~\citep{mimiccxrjpg}, which contains chest X-ray images. We pre-process the data and introduce new benchmark results for two tasks~\citep{benchhmark}: in-hospital mortality prediction and phenotype classification. The results show that the model's performance remains robust across uni-modal samples and improves for paired multi-modal samples. The model achieves state-of-the-art results without imposing any assumptions on correlation between modalities.
\item Considering the lack of multi-modal learning benchmarks in healthcare, we release our data pre-processing and benchmark code to allow reproducibility of the results and enable the evaluation of competing models in the future. The code can be found at: \url{https://github.com/nyuad-cai/MedFuse}. An overview of the proposed work is shown in Figure~\ref{fig:overview-of-work}.
\end{itemize}

\subsection*{Generalizable Insights about Machine Learning in the Context of Healthcare} \vspace{-1mm}
State-of-the-art multi-modal fusion approaches typically investigate synchronous sources of information using natural datasets, such as audio, visual, and textual modalities. In healthcare, data is often sparse and heterogeneous and hence modalities are not always paired. Our work overcomes the challenge of missing data by proposing a flexible fusion approach that is agnostic to the modality-specific encoders. Therefore, it can be used for other types of input data, beyond chest X-ray images and clinical time-series data. It also highlights the value of processing a sequence of uni-modal representations, compared to the conventional concatenation strategy in joint fusion. Overall, the work highlights the promise of multi-modal fusion in healthcare to improve  performance in downstream tasks.

% This section is \emph{required}, must keep the above title, and should
% be the final part of your introduction.  In about one paragraph, or
% 2-4 bullet points, explain what we should \emph{learn} from reading
% this paper that might be relevant to other machine learning in health
% endeavors.

% For example, a work that simply applies a bunch of existing algorithms
% to a new domain may be useful clinically but doesn't increase our
% understanding of the machine learning and healthcare; if that study
% also investigates \emph{why} different approaches have different
% performance, that might get us excited!  A more theoretical machine
% learning work may be in how it enables a new kind of clinical study.
% \emph{Reviewers and readers will look to evaluate (a) the significance
%   of your claimed insights and (b) evidence you provide later in the
%   work of you achieving that contribution}

%   Conventional multi-label chest X-ray classification models are constrained by the availability of data and its annotations. Our work overcomes the challenge of collecting large-scale annotated datasets by leveraging the use of rich medical literature, since it is the main knowledge source for all the discovered diseases by the medical community. This highlights the role of multi-modal learning in healthcare applications. Although we focus on chest X-rays, the network design can be potentially generalized to any medical imaging task, since the semantic encoder is task-agnostic. Improving the diagnosis of unseen diseases at inference has the potential to save patient lives.

\vspace{-0.5mm}
\section{Related Work}

Routine clinical practice produces large amounts of data from different sources (i.e. modalities), including medical images, laboratory test results, measurements of vital signs, and clinical notes~\citep{asri2015big}. Advances in deep learning have enabled building predictive models using subsets of modalities, typically clinical time-series data~\citep{shickel2017deep} and medical images~\citep{litjens2017survey}. Here, we provide an overview of related work on multi-modal fusion in healthcare using imaging and non-imaging data.

\subsection{Multi-modal Learning}
Multi-modal learning has been widely explored for jointly learning representations of multiple modalities~\citep{baltruvsaitis2018multimodal}. Example tasks include visual grounding~\citep{chen2021endtoend}, language grounding through visual cues~\citep{zhang2021explainable}, action recognition~\citep{chen2015utd}, video classification~\citep{nagrani2021attention}, image captioning~\citep{yu2019multimodal}, or visual-question answering~\citep{zellers2021merlot}. Since machine learning studies typically investigate different combinations of audio, visual, and textual modalities, many of the existing methods are driven by the assumption that the modalities share intrinsic and structural information. This is not always true for heterogeneous data in healthcare. Hence, due consideration should be given to learning with multiple medical data modalities, since conventional assumptions for non-medical data are not necessarily applicable.

%For example, variations in respiration rate can be attributed to different underlying medical conditions~\citep{Respiratory_Rate}. Based on the symptoms and the clinician's recommendations, different types of data can be collected to reach a final diagnosis.

\subsection{Multi-modal Fusion with Medical Images}
There is an increasing interest in advancing the fusion of multi-modal medical images~\citep{imaging_fusion}. The images usually represent different views of the same organ or lesion of interest, acquired using one or more sensors, whereby the images share the same set of labels. Proposed methods mainly focus on pixel-level fusion of complementary views acquired through multiple sensors to obtain a unified composite representation of the raw images~\citep{img_fusion, JAMES20144}. Various feature- and prediction-level fusion approaches were proposed for improved classification~\citep{Therapy_Response, Breast_Cancer, cancersubtypes} or segmentation performance~\citep{imaging_fusion}. Since textual reports are a natural byproduct of radiology exams, they were also used as additional modalities for tasks like visual-question answering~\citep{imag_reports, Sharma2021}, report generation~\citep{radiology}, or zero-shot image classification~\citep{gzsl, MVSE}.

\subsection{Multi-modal Fusion with Clinical Data and Medical Images} 
% Summary of applications pertaining to EHR + medical imaging 
Several studies investigated the fusion of medical images and clinical data extracted from the patient's Electronic Health Records (EHR) for various applications~\citep{Huang2020_survey}. For example, a stream of work covers tasks pertaining to cancer, such as recurrence prediction~\citep{cancer_recurrence}, lesion detection \citep{cancers_diag}, or patient survival prediction~\citep{cancer_survival}. Other tasks include detection of pulmonary embolism~\citep{Huang2020}, predicting the progression of Alzheimer’s disease~\citep{MildInt}, diagnosis of neurological disease~\citep{9534148}, or diagnosis of cervical dysplasia~\citep{cervical}. While these studies highlight the impact of using multiple data modalities on downstream performance, many curate datasets for specific tasks and share the assumption that the images and selected clinical features are paired.

Some studies specifically focused on the integration of clinical data and chest X-ray images. For example, the integration of the two modalities showed a favorable impact on the predictive performance in prognostication tasks among patients with COVID-19~\citep{Shamout2021, Jiao2021}. Some studies jointly refine a common latent representation after aggregating encoded features of each modality~\citep{Cardiomegaly_ehr_cxr, Jiao2021}, while others combine predictions computed by each modality through weighted averaging (i.e., late fusion)~\citep{Shamout2021, Jiao2021}. While late fusion enables the computation of predictions even for incomplete samples, it requires that the two modalities are assigned the same labels, which is not always feasible. Closely related to our work is that of \cite{hayat2021dynamic}, where they propose a dynamic training approach for partially paired clinical time-series data and chest X-ray images for the task of phenotype classification. However, their method is not scalable since it incorporates an additional classifier (and prediction) for every possible combination of input modalities.

\begin{figure*}[t!]
    \centering
    \includegraphics[width=.95\textwidth]{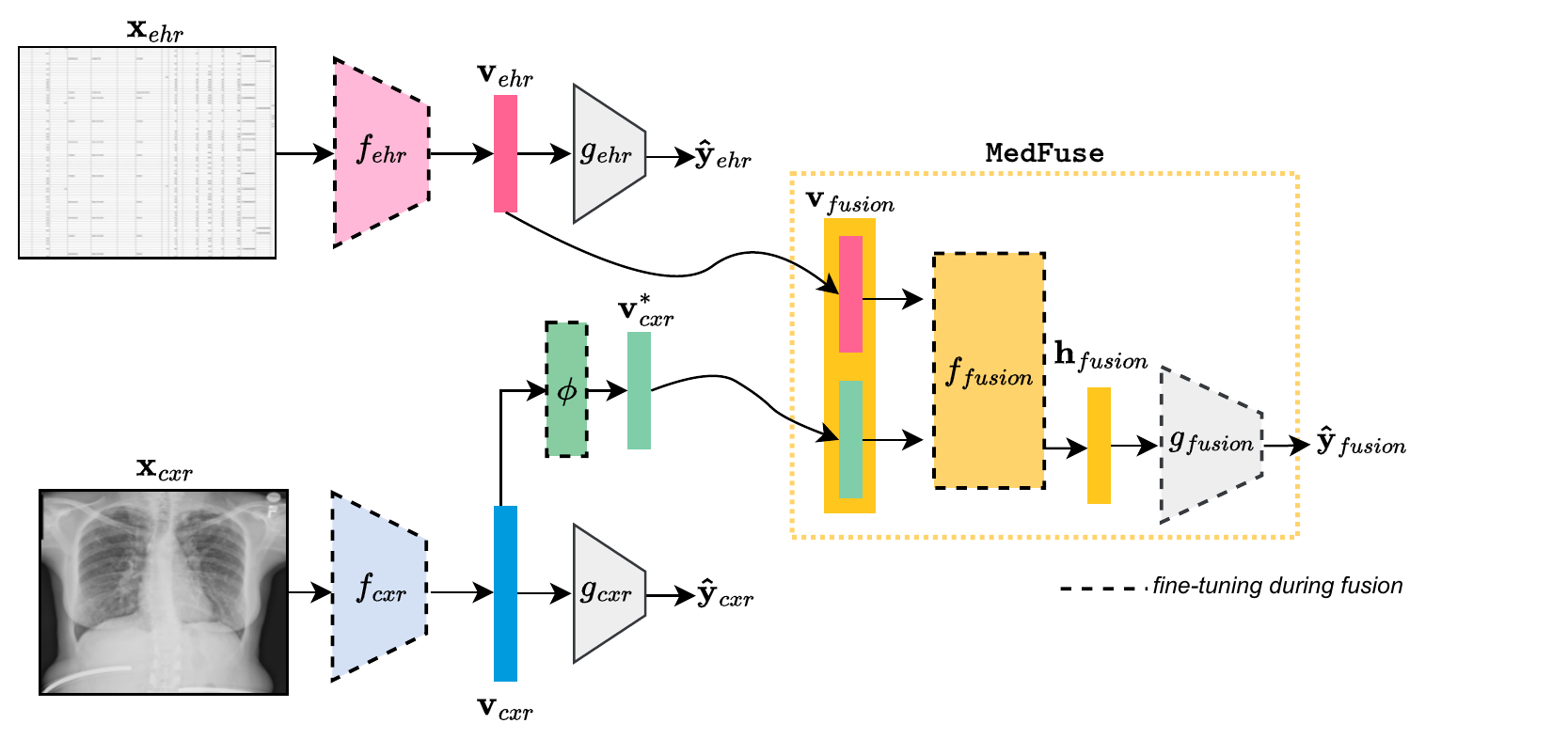}
    \vspace{-5mm}
    \caption{\small\textbf{Overview of network with \texttt{MedFuse} module.} First, we pre-train the modality-specific encoders and classifiers independently for each input modality. Specifically, we train $f_{ehr}$ and $g_{ehr}$ using the clinical time-series data and $f_{cxr}$ and $g_{cxr}$ using the chest X-ray images. Next, we project the chest X-ray latent representation $\mathbf{v}_{cxr}$ to $\mathbf{v}^*_{cxr}$, in order to match the dimension of $\mathbf{v}_{ehr}$. We pass $\mathbf{v}_{ehr}$ and $\mathbf{v}^*_{cxr}$ as an input sequence to the LSTM-based $f_{fusion}$, and we classify its last hidden state $\mathbf{h}_{fusion}$ to compute the overall prediction $\mathbf{\hat{y}}_{fusion}$. $f_{fusion}$, $f_{ehr}$, $f_{cxr}$, $g_{fusion}$, and $\phi$ are fine-tuned together for fusion.\vspace{-5mm} }
    \label{fig:main_fig}
    \vspace{-4mm}
\end{figure*}

\vspace{-0.5mm}
\section{Methodology}
\label{sec:method}

We define a two stage-approach (i) to learn modality-specific perceptual models to extract the latent features (Section~\ref{encoders}), and (ii) integrate these features through a joint multi-modal fusion module, \texttt{MedFuse} (Section~\ref{fusion}). The overall architecture is shown in Figure~\ref{fig:main_fig}. Without loss of generality, we focus here on two modalities only and denote the clinical time-series data as $ehr$ and the chest X-ray images as $cxr$ when defining the methodology.

\subsection{Modality-specific Encoders} \label{encoders}
One of the main sources of heterogeneity in healthcare is the varying dimensionality of the input modalities, which makes it challenging to develop a unified encoder for all input modalities. Another difference is the target space, since we do not assume that the modalities must be assigned the same set of labels. Hence, we first define modality-specific encoders as follows.

For a given instance, let $\mathbf{x}_{ehr}\in \mathbb{R}^{t\times d}$ represent the clinical time-series data associated with ground-truth labels $\textbf{y}_{ehr}$, where $t$ is the number of time steps and $d$ is the number of features derived from the clinical variables. We implement the encoder, $f_{ehr}$, for the clinical time-series modality as two stacked layers of an LSTM network~\citep{hochreiter1997long} with a dropout layer. We compute a latent feature representation $\mathbf{v}_{ehr} \in \mathbb{R}^m$ consisting of the last hidden state of the stacked LSTM, where $m=256$. We then apply a classifier, $g_{ehr}$, to compute the predictions, such that $\hat{\mathbf{y}}_{ehr} = g_{ehr}(\mathbf{v}_{ehr})$. To fine-tune the encoder, we optimize the following loss:
\begin{equation}
    \mathbb{L}_{ehr}(\mathbf{y}_{ehr}, \mathbf{\hat{y}}_{ehr}) = BCE(\mathbf{y}_{ehr}, \mathbf{\hat{y}}_{ehr}),
    \label{eqn:learning_obj_ehr}
\end{equation}
where $BCE$ is the Binary Cross-Entropy loss. 

Let $\mathbf{x}_{cxr} \in \mathbb{R}^{w\times h \times c}$ represent the chest X-ray image belonging to the same instance associated with the ground-truth labels $\textbf{y}_{cxr}$, where $w$ is the width dimension, $h$ is the height dimension, and $c$ is the number of channels. In all of our experiments, $h=224$, $w=224$, and $c=3$, as we replicate each image across three channels. We implement the encoder, $f_{cxr}$, as a ResNet-34~\citep{he2016deep} to compute $\mathbf{v}_{cxr} \in \mathbb{R}^n$, which is the feature representation after the average pooling layer of the convolutional network where $n=512$. Similarly, we then apply a classifier, $g_{cxr}$,  to compute the predictions, such that $\hat{\mathbf{y}}_{cxr} = g_{cxr}(\mathbf{v}_{cxr})$ and optimize the following loss to fine-tune the encoder:
\begin{equation}
    \mathbb{L}_{cxr}(\mathbf{y}_{cxr}, \mathbf{\hat{y}}_{cxr}) = BCE(\mathbf{y}_{cxr}, \mathbf{\hat{y}}_{cxr}). 
    \label{eqn:learning_obj_cxr}
\end{equation}

\noindent The encoders can hence be independently pre-trained using their respective labels and losses.

% such that $\mathbf{f_{cxr}(x_{cxr})}: \mathbf{X^{cxr}} \xrightarrow[]{} \mathbf{F^{cxr}}$, where $\mathbf{x_{cxr}}$ is the input sample from distribution $\mathbf{X^{cxr}} \in \mathbb{R}^{w\times h \times c}$.  The features are mapped to a latent space $\mathbf{F^{cxr}} \in \mathbb{R}^{d_{cxr}^{latent}}$, where ${d_{cxr}^{latent}}$ is the dimension of latent feature space. 

% such that $\mathbf{f_{ehr}(x_{ehr})}: \mathbf{X^{ehr}} \xrightarrow[]{} \mathbf{F^{ehr}}$, where $\mathbf{x_{ehr}}$ is the input sample from distribution $\mathbf{X^{ehr}} \in \mathbb{R}^{e \times d_{ehr}^{input}}$. Here, $d_{ehr}^{input}$ represents the number of input clinical variables and $e$ represents the number of time steps. The encoded features belong to the latent space $\mathbf{F^{ehr}} \in \mathbb{R}^{d_{ehr}^{latent}}$, where $d_{ehr}^{latent}$ is the latent space dimension. 

%  We define classification heads with activation for each encoder, such that $\mathbf{g_{ehr}}: \mathbf{F^{ehr}} \xrightarrow[]{} \mathbf{Y^{ehr}}$ for the clinical data, and $\mathbf{g_{cxr}}: \mathbf{F^{cxr}} \xrightarrow[]{} \mathbf{Y^{cxr}}$ for the chest X-ray images. Without loss of generality, the overall objective for each encoder is:
% \begin{equation}
%     \min_{\mathbf{f_i, g_i}} \mathbb{L}\mathbf{(x_i, y_i)} = \mathbb{L}(\mathbf{g_i(f_i(x_i)}), \mathbf{y_i}),
%     \label{eqn:learning_obj}
% % \end{equation}
% where $\mathbf{i} \in \{ \mathbf{ehr, cxr}\}$. For the clinical time-series data, we use the last hidden state of the encoder output as the input to the linear classification layer.

\subsection{The \texttt{MedFuse} Module}
\label{fusion}
To fuse the modalities, we first dismiss the classifiers, $g_{ehr}$ and $g_{cxr}$, and keep the the pre-trained modality-specific encoders, $f_{ehr}$ and $f_{cxr}$. Since the latent space dimensions of the two modalities are different, we use a projection layer, $\mathbf{\phi}$, that projects $\mathbf{v}_{cxr}$ to the same dimensionality as $\mathbf{v}_{ehr}$:
\begin{equation}
    \mathbf{v}_{cxr}^* =  { \phi(\mathbf{v}_{cxr})}
    \label{eqn:feats_proj}
\end{equation}
such that $\mathbf{v_{cxr}^*}\in \mathbb{R}^m$. We then create an input sequence consisting of the the uni-modal feature representations of the sample:
\begin{equation}
    \mathbf{v}_{fusion} = [\mathbf{v}_{ehr}, \mathbf{v}_{cxr}^*].
    \label{eqn:feats_concat}
\end{equation}
We parameterize a multi-modal fusion network, $f_{fusion}$, as a single LSTM layer with input dimension of 256 and a hidden dimension of 512, that aggregates the multi-modal sequence through recurrence. The motivation for using an LSTM is two-fold. First, it follows the intuition of decision-making, where clinicians examine information from each modality sequentially, or one at a time. This allows the LSTM module to initially learn from $\mathbf{v}_{ehr}$, and then update its internal state using information in $\mathbf{v}_{cxr}^*$. Second, it can handle input sequences of variable number of modalities, so it inherently deals with missing modalities. In the case that the chest X-ray image is missing during training or inference, the network processes a single-element sequence, $[\mathbf{v}_{ehr}]$.

The last hidden state, $\textbf{h}_{fusion}$, of $f_{fusion}$ is then processed using a classifier $g_{fusion}$ that computes the final fusion predictions, such that $\mathbf{\hat{y}}_{fusion}=g_{fusion}(\mathbf{h}_{fusion})$. We jointly train the encoders $f_{ehr}$ and $f_{cxr}$, the projection layer $\phi$, the fusion module $f_{fusion}$, and the classifier $g_{fusion}$, by optimizing the following loss:
\begin{equation}
    \mathbb{L}_{fusion}(\mathbf{y}_{fusion}, \mathbf{\hat{y}}_{fusion}) = BCE(\mathbf{y}_{fusion}, \mathbf{\hat{y}}_{fusion}),
    \label{eqn:learning_obj_fusion}
\end{equation}
where $\textbf{y}_{fusion}=\textbf{y}_{ehr}$, since we assume that the clinical time-series data modality is the base modality associated with the prediction task of interest, and is always present during training and inference. All classifiers $g_{ehr}$, $g_{cxr}$, and $g_{fusion}$ consist of a single linear layer followed by sigmoid activation.

\vspace{-0.5mm}
\section{Experiments}
\label{sec:exp}
\subsection{Datasets and Benchmark Tasks} For our experiments, we extract the clinical time-series data from MIMIC-IV~\citep{mimic4} along with the associated chest X-ray images in MIMIC-CXR~\citep{mimiccxrjpg}. 
%\subsubsection{Benchmark Tasks}
Here we describe the two tasks and provide more details on each:
\vspace{-0.5mm}
\begin{itemize}\setlength\itemsep{-0.5em}
    \item \textbf{Phenotype classification:} The goal of this multi-label classification task is to predict whether a set of 25 chronic, mixed, and acute care conditions are assigned to a patient in a given ICU stay. For a given instance, $\mathbf{x}_{ehr}$ contains clinical time-series data collected during the entire ICU record, and $\mathbf{y}_{ehr}$ is a vector of 25 binary phenotype labels. We link each instance with the last chest X-ray image collected during the same ICU stay. MIMIC-III contains International Classification of Diseases (ICD) version 9 (ICD-9) codes, whereas MIMIC-IV contains both ICD-9 and ICD-10. In the original benchmark paper~\citep{benchhmark}, the 25 phenotype labels were initially defined using the Clinical Classifications Software (CCS) for ICD-9~\citep{ccs_9}. Since ICD-9 and ICD-10 codes are aggregated to different CCS categories, we mapped all ICD-10 codes to ICD-9 using the guidelines provided by the Centers for Medicare \& Medicaid Services\footnote{Centers for Medicare \& Medicaid Services, \url{https://www.cms.gov/Medicare/Coding/ICD10/2018-ICD-10-CM-and-GEMs}}, and then map them to CCS categories. We evaluate this task using the Area Under the Receiver Operating Characteristic (AUROC) curve and the Area Under the Precision Recall curve (AUPRC).
     \item \textbf{In-hospital mortality prediction:} The goal of this binary classification task is to predict in-hospital mortality after the first 48 hours spent in the ICU. Hence, for a given instance, $\mathbf{x}_{ehr}$ contains clinical time-series data collected during the first 48 hours of the ICU record, and $\mathbf{y}_{ehr}$ is a binary label indicating in-hospital mortality. Since the task requires a minimum of 48 hours, we exclude ICU stays that are shorter than 48 hours. Here, we pair each instance with the last chest X-ray image collected during the ICU stay. We evaluate this task using AUROC and AUPRC.
    %  \item \textbf{Decompensation prediction:} The goal of this binary classification task is to predict mortality in the next 24 hours at each hour in the ICU. In this case, for a given instance, $\mathbf{x}_{ehr}$ contains clinical time-series data collected up to the respective hour, and $\mathbf{y}_{ehr}$ is a binary label indicating mortality in the next 24 hours. We pair each instance with the last chest X-ray image collected prior to the time of prediction. We evaluate this task using AUROC and AUPRC.
    %  \item \textbf{Length of Stay (LOS) prediction:} The goal of this multi-class classification task is to predict remaining length of stay at each hour in the ICU. For a given instance, $\mathbf{x}_{ehr}$ contains clinical time-series data collected up to the respective hour, and $\mathbf{y}_{ehr}$ is a one-of-ten hot binary vector. The classes are: less than a day, one class per each day of the first week, greater than one week but less than two, and greater than two weeks. We pair each instance with the last chest X-ray image collected prior to the respective time of prediction. We use Cohen’s linear weighted kappa score as the evaluation metric.
\end{itemize}

%To link the clinical time-series data with chest X-ray images, we follow a similar approach to that described in recent work for the phenotyping task~\citep{hayat2021dynamic}.
% \subsection{Data Extraction} 
\subsubsection{Pre-processing of Clinical Time-series Data}
We modified the extraction and data pre-possessing pipeline of \cite{benchhmark}, which was originally implemented in TensorFlow \citep{tensorflow2015}, and introduce a new version for MIMIC-IV using Pytorch \citep{NEURIPS2019_9015}. To make a fair comparison and illustrate the efficacy of multi-modal learning, we use the same set of 17 clinical variables. Amongst those, five are categorical (capillary refill rate, Glasgow coma scale eye opening, Glasgow coma scale motor response, Glasgow coma scale verbal response, and Glasgow coma scale total) and 12 are continuous (diastolic blood pressure, fraction of inspired oxygen, glucose,  heart rate, height, mean blood pressure, oxygen saturation, respiratory rate, systolic blood pressure, temperature, weight, and pH). For all the tasks, we regularly sample the input every two hours, discretize and standardize the clinical variables to obtain the input for $f_{ehr}$ as in previous work~\citep{benchhmark}. After data pre-processing and one-hot encoding of the categorical features, we obtain a vector representation of size 76 at each time-step of the clinical time-series data, such that for a given instance, $\mathbf{x}_{ehr}\in\mathbb{R}^{t\times76}$ and $t$ depends on the instance and task.

%We also use International Classification of Diseases version 9 (ICD-9).

\subsubsection{Data Splits}
Using the patient identifier of the clinical time-series data, we randomly split the dataset into 70\% for training, 10\% for validation, and 20\% for test set, as shown in Figure~\ref{fig:overview-of-work}. We report final results on the test sets and compute 95\% confidence intervals with 1000 iterations via the bootstrap method~\citep{efron1994introduction}. Here, we denote the clinical time-series data as $\mathbf{EHR}$ and the chest X-ray images as $\mathbf{CXR}$.  $(\mathbf{EHR}+\mathbf{CXR})_\mathbf{PARTIAL}$ contains paired and partially paired samples (i.e. samples where chest X-ray is missing). $(\mathbf{EHR}+\mathbf{CXR})_\mathbf{PAIRED}$ contains data samples where both modalities are present. %We introduce additional notation to indicate subsets of those datasets: $\textbf{EHR}_\mathbf{PARTIAL}$ contains all clinical time-series partially paired with chest X-ray images, and $\mathbf{PAIRED_{EHR}}$ only contains clinical time-series data within $\mathbf{PAIRED}$.
For example, the $(\mathbf{EHR}+\mathbf{CXR})_\mathbf{PARTIAL}$ training set for patient phenotyping contains 7756 samples associated with chest X-rays amongst 42628 samples.  %The same datasets were used for training and evaluation of all baseline models. 

We extract from MIMIC-CXR chest X-ray images and split them based on a random patient split. We then transfer images from the training set to either the validation or test set, in case are were associated with patients in the validation or test splits of the clinical time-series data. This procedure resulted with 325188 images in the training set, 15282 images in the validation set, and 36625 images in the test set. We define  $\mathbf{y}_{cxr}$ as a vector of 14 binary radiology labels extracted from radiology reports through CheXpert~\citep{irvin2019chexpert}. We denote this uni-modal dataset as $\mathbf{CXR}_{\mathbf{UNI}}$ and it is fixed across all tasks.  We introduce an additional notation for $\mathbf{CXR}_{\mathbf{PAIRED}}$, which includes only chest X-ray images within $(\mathbf{EHR}+\mathbf{CXR})_\mathbf{PAIRED}$, and $\mathbf{EHR}_{\mathbf{PARTIAL}}$, which includes only clinical time-series data within $(\mathbf{EHR}+\mathbf{CXR})_\mathbf{PARTIAL}$.

\subsection{Training Strategy with the \texttt{MedFuse} Module}
The training strategy consists of two steps: pre-training of the modality-specific encoders followed by jointly fine-tuning the encoders and fusion module. During the pre-training stage, we train the image encoder using the full uni-modal training dataset $\mathbf{CXR}_{\mathbf{UNI}}$ with the 14 radiology labels. We also pre-train the clinical time-series data encoder for each task independently using the training sets $\mathbf{EHR}_{\mathbf{PARTIAL}}$, since each task is associated with its own set of inputs and labels. After pre-training the modality-specific encoders, we discard the uni-modal classifiers and fine-tune the encoders, projection layer, and \texttt{MedFuse} using $(\mathbf{EHR+CXR})_{\mathbf{PARTIAL}}$. We compare this training strategy to fine-tuning the fusion module with randomly initialized feature extractors.

% After tuning the learning rate with the hyperparameter experiments, we fix the learning rate and select a specific percentage of training samples with a missing mdoality. 
% Since there is a high imbalance between the number of uni-modal and paired multi-modal samples in $\mathbf{EHR}_{\mathbf{PARTIAL}}$, we include all of the paired multi-modal samples and select a specific percentage of training samples with a missing modality during each fine-tuning epoch. The optimal 

\begin{figure}
    \centering
    \includegraphics[width=1.0\textwidth]{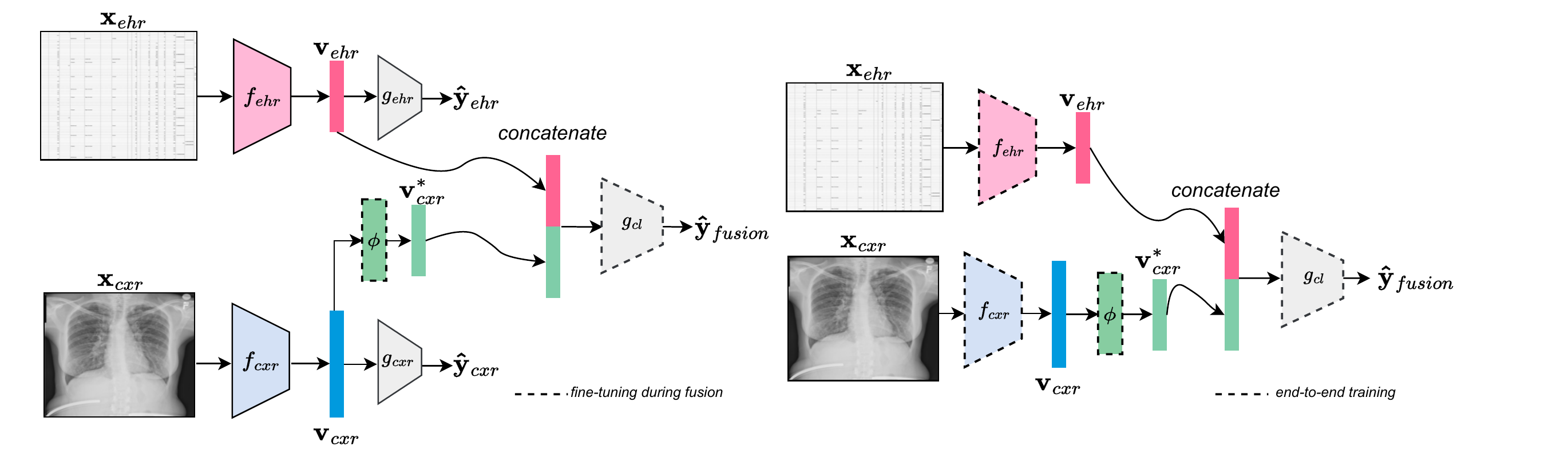}
    \caption{\small\textbf{Architecture of early and joint fusion baselines.} In early fusion (left), the encoders are first pre-trained. Then, we freeze them and fine-tune the projection layer and fusion classification module. In joint fusion (right), the encoders and classification module are randomly initialized and trained end-to-end.\vspace{-10mm}}
    \label{fig:baselines}
\end{figure}

\subsection{Baseline Models}
We compare the performance of our proposed multi-modal approach to several existing baselines:
\vspace{-2mm}
% [itemsep=-4pt, topsep=-3pt]
\begin{itemize}\setlength\itemsep{-0.3em}
\item \textbf{Early fusion:} The vanilla early fusion approach commonly used in recent work~\citep{Huang2020_survey} (Figure~\ref{fig:baselines} (left)) assumes the presence of paired data modalities during training and inference. We train two versions. In the first version, we pre-train modality-specific networks independently: $f_{cxr}$ and $g_{cxr}$ with the $\mathbf{CXR}_\mathbf{PAIRED}$ training set, and $f_{ehr}$ and $g_{ehr}$ with the $\mathbf{EHR}_\mathbf{PAIRED}$ training set. We then freeze the encoders $f_{cxr}$ and $f_{ehr}$, concatenate their latent feature representations, and fine-tune a projection layer and a fully connected classification network, denoted as $g_{cl}$ using the $(\mathbf{EHR}+\mathbf{CXR})_\mathbf{PAIRED}$ training set. In the second version, we use the $(\mathbf{EHR}+\mathbf{CXR})_\mathbf{PARTIAL}$ training set for fine-tuning the projection layer and $g_{cl}$. Inspired by \cite{kyono2021miracle}, we learn a vector to substitute for missing chest X-ray images.  

\item \textbf{Joint fusion:} In this setting, we train a network end-to-end including the modality-specific encoders ($f_{cxr}$ and $f_{ehr}$) and a classification network applied to the concatenated latent representations of the two encoders (Figure~\ref{fig:baselines} (right)). We train two versions. In the first version, we train a randomly initialized network end-to-end using $(\mathbf{EHR}+\mathbf{CXR})_\mathbf{PAIRED}$. In the second version, we train a randomly initialized network end-to-end using $(\mathbf{EHR}+\mathbf{CXR})_\mathbf{PARTIAL}$ with a learnable vector to substitute for any missing chest X-ray images. %We tune the learning rate during hyperparameter tuning over 10 runs for each version and choose the model checkpoint that achieves the best AUROC on the respective validation set. 

\item \textbf{Multi-modal Transfer Module (MMTM):} Originally proposed by~\cite{vaezi20mmtm}, this approach also assumes paired input data. We apply an MMTM module after the first LSTM layer in the clinical time-series modality, and either the third or the fourth ResNet layer. We train a randomly initialized network with the MMTM module end-to-end using the $(\mathbf{EHR}+\mathbf{CXR})_\mathbf{PAIRED}$ training set, and closely follow the training strategy described in the original paper.\footnote{https://github.com/haamoon/mmtm} %We tune the learning rate during hyperparameter tuning over 10 runs for each of the two settings, and choose the model checkpoint and setting based on the best results on the respective validation set. 

\item \textbf{Dynamic Affine Feature Map Transform (DAFT):} Also requiring paired input data, we use the general purpose DAFT module~\citep{daft} to rescale and shift the feature representations after the first LSTM layer using the chest X-ray representation computed either through the third or fourth layer of ResNet. Similarly, we use $(\mathbf{EHR}+\mathbf{CXR})_\mathbf{PAIRED}$, and follow the training approach in the original work's respository.\footnote{https://github.com/ai-med/DAFT/} %We tune the learning rate during hyperparameter tuning over 10 runs for each of the two settings, and choose the model checkpoint and setting based on the best results on the respective validation set. 
\end{itemize} 

\noindent We also compare it with a uni-modal two-layer LSTM network trained with clinical time-series data only, and the method proposed by~\cite{hayat2021dynamic} (Unified) trained with $(\mathbf{EHR}+\mathbf{CXR})_\mathbf{PARTIAL}$. 

% Since the design of these conventional approaches requires both modalities to be present at the fusion stage, we compare them to \texttt{MedFuse} on a the test set that only contains paired samples ($\mathbf{PAIRED}$). 

% \subsection{Implementations Details}
% % \subsection{data extraction and linking} We extract the dataset 
% In this section, we describe important technical implementation details.

% \subsubsection{Image augmentations}

\begin{table*}[t!]
    \centering
    \caption{\small\textbf{Performance results in the uni-modal vs multi-modal setting.} Here, we compare the stacked LSTM network for the clinical time-series data only, with our network using \texttt{MedFuse}. In the first four rows, we summarize the AUROC and AUPRC results in the paired setting with uni-modal ($\mathbf{EHR}_{\mathbf{PAIRED}}$) and multi-modal data ($(\mathbf{EHR}+\mathbf{CXR})_{\mathbf{PAIRED}}$). In the last two rows, we show the results on the partially paired test set, with the uni-modal subset ($\mathbf{EHR}_{\mathbf{PARTIAL}}$) and multi-modal data ($(\mathbf{EHR}+\mathbf{CXR})_{\mathbf{PAIRED}}$). All results shown below are for \texttt{MedFuse} (OPTIMAL). Best results are shown in bold.} \vspace{-1mm}
     \resizebox{1.0\textwidth}{!}{\begin{tabular}{l c c c c| c c c} \toprule
     & \multicolumn{ 2}{c}{\textbf{Modalities}} &  \multicolumn{ 2}{c}{\textbf{Phenotyping}} &  \multicolumn{ 2}{c}{\textbf{In-hospital mortality}} \\
     \midrule
    
    \textbf{Model} & \textbf{Training set} & \textbf{Test set}  & \textbf{AUROC} & \textbf{AUPRC} & \textbf{AUROC} & \textbf{AUPRC} \\
    \midrule
       LSTM &   $\mathbf{EHR}_{\mathbf{PAIRED}}$  & $\mathbf{EHR}_{\mathbf{PAIRED}}$ & 0.716 & 0.407 & 0.818 & 0.460   \\
        &    &  & (0.688, 0.743) & (0.367, 0.453) &  (0.787, 0.845) &  (0.395, 0.535)  \\
       
       LSTM &   $\mathbf{EHR}_{\mathbf{PARTIAL}}$  & $\mathbf{EHR}_{\mathbf{PAIRED}}$ & 0.746  & 0.453  & 0.825  & 0.500  \\
        &     &  &  (0.720, 0.772) & (0.409, 0.502) &  (0.793, 0.852) &  (0.428, 0.576)  \\
       
       \texttt{MedFuse} & $\mathbf{(EHR+CXR)}_{\mathbf{PARTIAL}}$  & $\mathbf{EHR}_{\mathbf{PAIRED}}$  & 0.740  & 0.441  & 0.833  & 0.514  \\
       &   &   &  (0.713, 0.767) &  (0.398, 0.489) &  (0.802, 0.861) &  (0.443, 0.584) \\
        
     \texttt{MedFuse} & $\mathbf{(EHR+CXR)}_{\mathbf{PARTIAL}}$  & $\mathbf{(EHR+CXR)}_{\mathbf{PAIRED}}$  & \textbf{0.770}  & \textbf{0.481}  & \textbf{0.865}  & \textbf{0.594}  \\
         &   &   &  (0.745, 0.795) &  (0.436, 0.531) &  (0.837, 0.889) &  (0.526, 0.655) \\

        \hline
     LSTM &   $\mathbf{EHR}_{\mathbf{PARTIAL}}$  & $\mathbf{EHR}_{\mathbf{PARTIAL}}$  & 0.765   & 0.425  & 0.861   &0.522  \\
      &  &   &  (0.754, 0.777)  &  (0.404, 0.447) &(0.846, 0.876)  & (0.482, 0.564) \\
    %  \texttt{MedFuse} & $\mathbf{EHR}_{\mathbf{PARTIAL}}$  & $\mathbf{PAIRED}_{\mathbf{EHR}}$ & \\
        \texttt{MedFuse} & $\mathbf{(EHR+CXR)}_{\mathbf{PARTIAL}}$  & $\mathbf{(EHR+CXR)}_{\mathbf{PARTIAL}}$ & \textbf{0.768}  & \textbf{0.429} & \textbf{0.874}  &\textbf{0.567}  \\
        &   &  &  (0.756, 0.779) &  (0.408, 0.452)& (0.860, 0.888) & (0.529, 0.607) \\
       \bottomrule
    \end{tabular}}
    \label{tab:univsmulti} \vspace{-3mm}
\end{table*}

\subsection{Model Training and Selection}
We perform hyperparameter tuning over 10 runs for our proposed network with \texttt{MedFuse} and each of the baseline models and their different versions. In each run, we randomly sample a learning rate between $10^{-5}$ and $10^{-3}$, and then choose the model and learning rate that achieve the best AUROC on the respective validation set. For the baselines with architectural choices (i.e. MMTM and DAFT), we choose the architecture that achieves the best performance on the validation set, and report its results on the test set. We use the Adam optimizer~\citep{kingma2014adam} across all experiments with a batch size of 16. We set the maximum number of epochs to 50 and use early stopping if the validation AUROC does not improve for 15 epochs. We also apply image augmentations as described in Appendix~\ref{image-aug}.

%We also investigate the impact of learning with unpaired data samples with the \texttt{MedFuse} module. 

With the best learning rate chosen via hyperparameter tuning, we vary the percentage of samples with $\mathbf{EHR}$ only data in the $(\mathbf{EHR+CXR})_{\mathbf{PARTIAL}}$ training set, fine-tune \texttt{MedFuse} accordingly and evaluate it on the validation set. We select the best model based on the best AUROC performance on the $(\mathbf{EHR+CXR})_{\mathbf{PARTIAL}}$ validation set, and report its results on the test set. We denote this chosen model as \texttt{MedFuse} (OPTIMAL).

% We train the visual network for 100 epochs, while all other training run for 50 epochs.  We used a single NVIDIA RTX 6000 GPU. It takes around 36 hours to train the chest X-ray encoder, 4 hours to train the clinical data encoder for the phenotyping task, and 2 hours to train the clinical data encoder for the in-hospital mortality task. Fine-tuning the \texttt{MedFuse} module along with the encoders took around 5 hours for the phenotyping task and 3 hours for in-hospital mortality task. 

\begin{table*}[t!]
    \centering
    \caption{\small\textbf{Performance results on the ($\mathbf{EHR+CXR})_{\mathbf{PAIRED}}$ test set.} We show the AUROC and AUPRC results for our proposed approach with \texttt{MedFuse} and the baseline models. We include results for early and joint fusion when trained with either  ($\mathbf{EHR+CXR})_{\mathbf{PAIRED}}$ or ($\mathbf{EHR+CXR})_{\mathbf{PARTIAL}}$, where the latter uses a learnable vector in the case of a missing chest X-ray image. We also show results of our proposed approach when we fine-tune the fusion module with ($\mathbf{EHR+CXR})_{\mathbf{PARTIAL}}$ and randomly initialized encoders (RI) or pre-trained encoders (PT), and the best version of the latter when using the optimal number of uni-modal samples during fine-tuning (OPTIMAL).  Best results are shown in bold.} \vspace{-1mm}
     \resizebox{1.0\textwidth}{!}{\begin{tabular}{l c c | c  c c} \toprule
     \textbf{Task} & \multicolumn{ 2}{c}{\textbf{Phenotyping}} &  \multicolumn{ 2}{c}{\textbf{In-hospital mortality}} \\ \midrule
    \textbf{Method} & \textbf{AUROC} & \textbf{AUPRC} & \textbf{AUROC} & \textbf{AUPRC} \\
    \midrule
       Early ($\mathbf{EHR+CXR})_{\mathbf{PAIRED}}$ & 0.753 (0.726, 0.779) & 0.453 (0.411, 0.502) & 0.827 (0.801, 0.854) & 0.485 (0.417, 0.555) \\
       Early ($\mathbf{EHR+CXR})_{\mathbf{PARTIAL}}$ &0.739 (0.712, 0.766) &0.435 (0.393, 0.483) & 0.818 (0.788, 0.845) & 0.467 (0.402, 0.539) \\ 
       \hline 
       Joint ($\mathbf{EHR+CXR})_{\mathbf{PAIRED}}$ & 0.747 (0.720, 0.773) & 0.446 (0.404, 0.493) & 0.825 (0.798, 0.853) & 0.506 (0.436, 0.574)  \\
       Joint ($\mathbf{EHR+CXR})_{\mathbf{PARTIAL}}$ & 0.754 (0.727, 0.780) &0.458 (0.415, 0.506) & 0.819 (0.785, 0.850) & 0.479 (0.413, 0.552) \\ 
       \hline 
       MMTM \citep{vaezi20mmtm} & 0.734 (0.707, 0.761) &  0.428 (0.387, 0.476)  & 0.819 (0.788, 0.846) & 0.474 (0.402, 0.544) \\
       DAFT \citep{daft} & 0.737 (0.710, 0.764) & 0.434 (0.393, 0.482)  & 0.828 (0.799, 0.854)& 0.492 (0.427, 0.572)\\ 
       Unified \citep{hayat2021dynamic} & 0.765 (0.742, 0.794) & 0.461 (0.417, 0.511) & 0.835 (0.808, 0.861) & 0.495 (0.424, 0.567) \\ 

       \midrule
       \texttt{MedFuse} (RI) & 0.748 (0.721, 0.774)  & 0.452 (0.408, 0.501) & 0.817 (0.785, 0.846) & 0.471 (0.404, 0.545) \\  
       \texttt{MedFuse} (PT) &0.756 (0.729, 0.782) & 0.466 (0.420, 0.515) & 0.841 (0.813, 0.868) & 0.544 (0.477, 0.609) \\
       \texttt{MedFuse} (OPTIMAL) & \textbf{0.770} (0.745, 0.795) & \textbf{0.481} (0.436, 0.531) & \textbf{0.865} (0.837, 0.889) & \textbf{0.594} (0.526, 0.655) \\  
       \bottomrule
    %   \multicolumn{5}{l}{\footnotesize{$^*$Note that the original paper only provided the AUROC results for the phenotyping task.}} \\
    \end{tabular}}
    \label{tab:paired_res}
    \vspace{-4mm}
\end{table*}

\vspace{-0.5mm}
\section{Results}
\label{sec:res}
In this section, we describe the results for a number of experiments to provide insights on our proposed approach. The learning rates that achieved the best results are summarized in Appendix~\ref{hyperparameter} for all models. The results on the validation set in the experiments where we vary the percentage of uni-modal samples during training are shown in Appendix~\ref{unimodalpercentagetraining}. The optimal percentages are 10\% for in-hospital mortality prediction, and 20\% for phenotype classification. 

\vspace{-2mm}
\subsection{Performance Results in the Uni-modal \& Multi-modal Settings}
In Table~\ref{tab:univsmulti}, we compare our proposed approach to the uni-modal stacked LSTM. As expected, we first observe that the performance of the uni-modal LSTM improves on the $\mathbf{EHR}_{\mathbf{PAIRED}}$ test set, in terms of AUROC and AUPRC for both tasks, when using the larger $\mathbf{EHR}_{\mathbf{PARTIAL}}$ training set.  Our proposed approach using \texttt{MedFuse} achieves the best performance on the paired test set when the chest X-ray images are used during training and inference as an auxiliary modality (0.770 AUROC and 0.481 AUPRC for phenotype classification, and 0.865 AUROC and 0.594 AUPRC for in-hospital mortality). We note similar, but less significant trends, in the larger partially paired test set, which may be due to the fact that only 18.8\% and 26.2\% of samples are paired in the phenotyping and in-hospital mortality test sets, respectively.

\vspace{-2mm}
\subsection{Performance Results in the Paired Setting}
Since the baseline models were originally designed for paired input, we evaluate all models on the ($\mathbf{EHR+CXR})_{\mathbf{PAIRED}}$ test set as shown in Table~\ref{tab:paired_res}. First, we observe that early fusion and joint fusion perform comparably across both tasks when trained with ($\mathbf{EHR+CXR})_{\mathbf{PAIRED}}$, with early fusion achieving a slightly better performance in terms of AUROC. We also note that training early fusion using ($\mathbf{EHR+CXR})_{\mathbf{PARTIAL}}$ leads to a drop in AUROC and AUPRC across both tasks, while joint fusion only improves for phenotype classification. Second, we observe that the Unified approach by \cite{hayat2021dynamic} achieves the best performance amongst all baseline approaches, with 0.765 AUROC and 0.461 AUPRC for phenotype classification, and 0.835 AUROC and 0.495 AUPRC for in-hospital mortality prediction. Third, we observe that our proposed approach with \texttt{MedFuse} (OPTIMAL) achieves the best performance across both tasks, with 0.770 AUROC and 0.481 AUPRC for phenotype classification, and 0.865 AUROC and 0.594 AUPRC for in-hospital mortality prediction. We also performed an ablation study where we randomly dropped the chest X-ray modality in the paired test set. The results are shown in Appendix~\ref{unimodalpercPAIRED}. We also compared the use of substituting the missing modality with zeros or with a learnable vector for early and joint fusion and the results are shown in Appendix~\ref{missingtoken}. The two techniques perform comparably.

%  Moreover, we observe that  achieves the best performance across both tasks. It is interesting to note that AUPRC is significantly improved for in-hospital mortality task (0.519 AUPRC by the best baseline model compared to 0.597 AUPRC by \texttt{MedFuse}). This highlights the importance of the chest X-ray modality for the task. 

\begin{table*}[t!]
    \centering
    \caption{\small\textbf{Performance results on the ($\mathbf{EHR+CXR})_{\mathbf{PARTIAL}}$ test set.} We compare our proposed approach with \texttt{MedFuse} with early and joint fusion when trained with ($\mathbf{EHR+CXR})_{\mathbf{PARTIAL}}$, including samples with missing chest X-ray images (substituted with a learnable vector). All methods were trained with the full ($\mathbf{EHR+CXR})_{\mathbf{PARTIAL}}$ training set, except for \texttt{MedFuse} (OPTIMAL) which uses the optimal number of uni-modal samples during fine-tuning. Best results are shown in bold.} \vspace{-1mm}
     \resizebox{1.0\textwidth}{!}{\begin{tabular}{l c c| c c } \toprule
     Task & \multicolumn{ 2}{c}{\textbf{Phenotyping}} &  \multicolumn{ 2}{c}{\textbf{In-hospital mortality}} \\
     \midrule
    
    \textbf{Method}  & \textbf{AUROC} & \textbf{AUPRC} & \textbf{AUROC} & \textbf{AUPRC} \\
    \midrule
       
       Early & 0.748 (0.735, 0.760) &0.394 (0.374, 0.416) & 0.860 (0.850, 0.877) & \textbf{0.515} (0.477, 0.556)  \\
       Joint & 0.754 (0.742, 0.766)& 0.410 (0.389, 0.433) & 0.841 (0.823, 0.857) & 0.482 (0.442, 0.525)\\
       \texttt{MedFuse}  & \textbf{0.758} (0.745, 0.770) & \textbf{0.418} (0.396, 0.441) & \textbf{0.861} (0.845, 0.874) & 0.501 (0.462, 0.543)   \\
       \midrule
       \texttt{MedFuse} (OPTIMAL) & \textbf{0.768} (0.756, 0.779) & \textbf{0.429} (0.408, 0.452)& \textbf{0.874} (0.860, 0.888) &\textbf{0.567} (0.529, 0.607)  \\
       
       \bottomrule
    \end{tabular}}
    \label{tab:partialresults}
\end{table*}

% both uni and multi modal use full ehr data in the table 4
\begin{figure}[t!]
    \centering
    % \vspace{-1.5mm}
    \includegraphics[width=.49\textwidth]{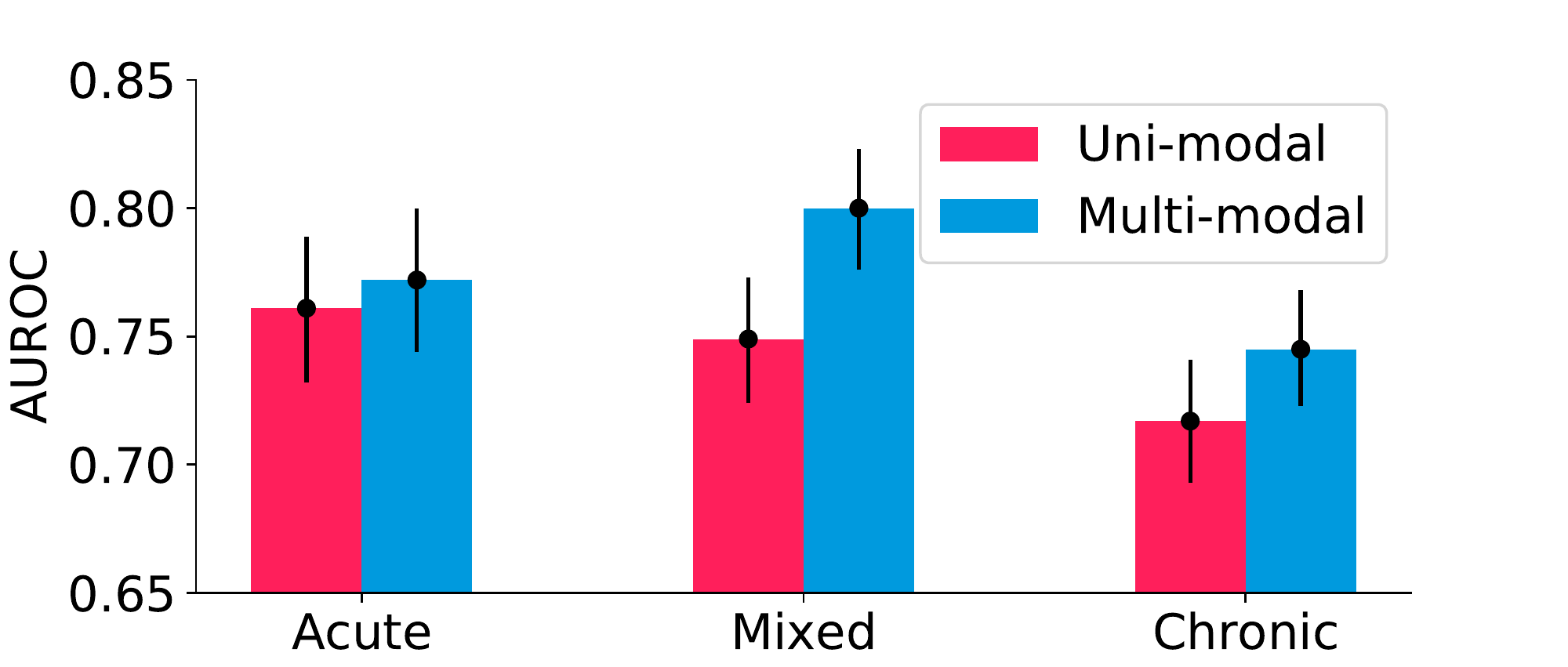}
    %\vspace{-0.5mm}
      \includegraphics[width=.49\textwidth]{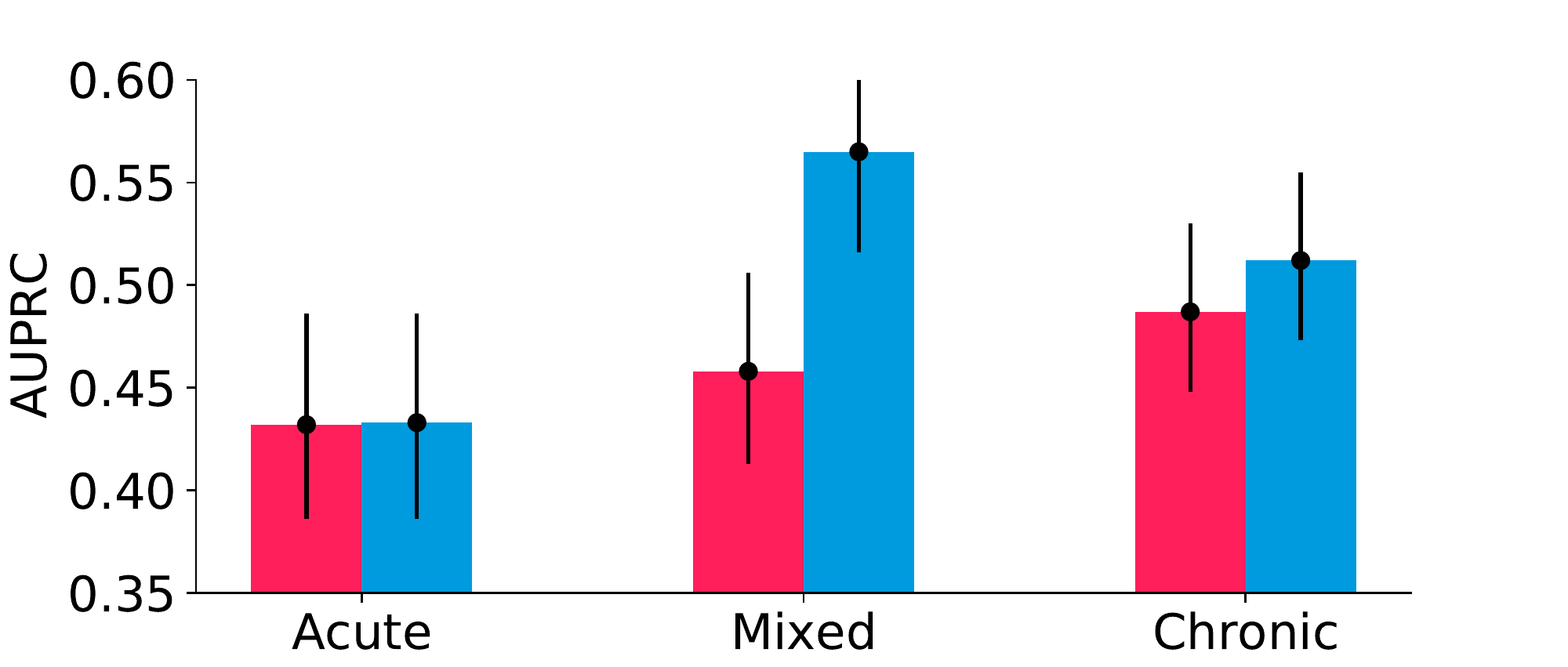}
    \caption{\small\textbf{Performance results across different subsets of labels for the phenotype classification task in the $(\mathbf{EHR}+\mathbf{CXR})_{\mathbf{PAIRED}}$ test set.} The multi-modal approach with \texttt{MedFuse} achieves the highest AUROC and AUPRC gains for the mixed conditions followed by chronic conditions, compared to the results achieved by the uni-modal stacked LSTM with $\mathbf{EHR}_{\mathbf{PAIRED}})$.} 
    \label{fig:types_bar}
    \vspace{-3.5mm}
\end{figure}

\subsection{Performance Results in the Partially Paired Setting} 
In Table~\ref{tab:partialresults}, we evaluate our proposed approach with \texttt{MedFuse} as well as early and joint fusion on the partially paired test set. Compared to early fusion, our proposed approach trained with the full $(\mathbf{EHR}+\mathbf{CXR})_{\mathbf{PARTIAL}}$ training set achieves a better performance for phenotype classification (0.758 compared to 0.748 AUROC and 0.418 compared to 0.394 AUPRC). It performs comparably with early fusion in the in-hospital mortality prediction task, although early fusion achieves a better AUPRC. Our approach outperforms joint fusion in the in-hospital mortality setting (0.861 compared to 0.841 AUROC and 0.501 compared to 0.482 AUPRC), and performs comparably for phenotype classification. Overall, \texttt{MedFuse} (OPTIMAL), fine-tuned with paired samples and only 10\% of uni-modal samples for in-hospital mortality prediction and 20\% of uni-modal samples for phenotype classification, achieves the best performance (0.768 AUROC and 0.429 AUPRC for phenotype classification and 0.874 AUROC and 0.567 AUPRC for inhospital mortality prediction). We also performed an ablation study where we varied the percentage of uni-modal samples in the partially paired setting. The results are shown in Appendix~\ref{unimodalpercPARTIAL}.

We also compared the performance of \texttt{MedFuse} to an ensemble consisting of (i) \texttt{MedFuse} for paired samples, and (ii) a uni-modal LSTM for samples with missing chest X-rays. While the results are comparable, as shown in Appendix~\ref{ensemble}, the results imply that an ensemble of strong models may be better suited for some tasks, such as phenotyping. This however requires the training of two models.

% In Table~\ref{tab:partialresults}, we perform the following comparisons: 

% \subsection{Impact of Learning with Missing Data} 
% In Table~\ref{tab:robustness}, we report the results when using different training sets to evaluate the impact of learning with missing modalities. As expected, training with all the samples of clinical time-series data (i.e., \textbf{ehr} (full)) performs better than training with the smaller subset that is associated with chest X-ray images (i.e., \textbf{ehr} (paired)) for both phenotyping (0.741 vs. 0.716 AUROC) and in-hospital mortality (0.821 vs. 0.814 AUROC). As shown in the third row, we observe that the incorporation of the additional modality achieves the best performance for both tasks in terms of AUROC and AUPRC, compared to training with a single modality. When considering the partially paired full test set (last two rows), our approach remains robust or improves the overall performance. In particular, we notice that the AUPRC increases from 0.518 to 0.552 for in-hospital mortality. TODO: discuss missing modality results with early and joint fusion (potentially move to appendix)

\begin{table*}[t!]
    \centering
    \caption{\small\textbf{Performance results across the different phenotype labels on $(\mathbf{EHR}+\mathbf{CXR})_{\mathbf{PAIRED}}$ test set, compared to the uni-modal stacked LSTM with $\mathbf{EHR}_{\mathbf{PAIRED}}$.} We report the performance results for the individual phenotypes using AUROC and AUPRC, and show the prevalence of labels in the $(\mathbf{EHR}+\mathbf{CXR})_{\mathbf{PARTIAL}}$ training set, and the $(\mathbf{EHR}+\mathbf{CXR})_{\mathbf{PAIRED}}$ test set. The labels and results in bold indicate that \texttt{MedFuse} achieved a performance improvement.}\vspace{-2mm}
     \resizebox{1.0\textwidth}{!}{\begin{tabular}{l c c c| c  c| c c c} \toprule
     & & \multicolumn{ 2}{c}{\textbf{Prevalence}} & \multicolumn{ 2}{c}{$\mathbf{EHR}_{\mathbf{PAIRED}}$} &  \multicolumn{ 2}{c}{$(\mathbf{EHR}+\mathbf{CXR})_{\mathbf{PAIRED}}$} &  \\
     \midrule
    
    \textbf{Phenotype} & \textbf{Type} & \textbf{Train} & \textbf{Test} & \textbf{AUROC} & \textbf{AUPRC} & \textbf{AUROC} & \textbf{AUPRC} \\
    \midrule
    % mixed acute chronic
    \textbf{Acute and unspecified renal failure} & acute & 0.269 & 0.321 & 0.780 (0.759, 0.800) & 0.614 (0.573, 0.655) & 0.\textbf{782} (0.760, 0.802) & \textbf{0.618} (0.579, 0.661)
 \\ 
    Acute cerebrovascular disease & acute& 0.056& 0.078 &0.903 (0.878, 0.929) & 0.501 (0.426, 0.591)  &0.888 (0.859, 0.915) & 0.496 (0.420, 0.578)\\ 
    \textbf{Acute myocardial infarction} & acute& 0.074& 0.093 & 0.729 (0.694, 0.765) & 0.267 (0.218, 0.329)  & \textbf{0.766} (0.732, 0.798) & \textbf{0.297} (0.237, 0.361)\\ 
    \textbf{Cardiac dysrhythmias} & mixed & 0.326 & 0.379 & 0.664 (0.640, 0.687) & 0.552 (0.517, 0.590)  & \textbf{0.708} (0.686, 0.730) & \textbf{0.581} (0.543, 0.618) \\ 
    \textbf{Chronic kidney disease} & chronic & 0.206& 0.240 & 0.748 (0.727, 0.771) & 0.457 (0.419, 0.505) & \textbf{0.768} (0.747, 0.789) & \textbf{0.485} (0.445, 0.533)\\ 
    \textbf{Chronic obstructive pulmonary disease} & chronic& 0.143 & 0.148 & 0.673 (0.640, 0.703) & 0.272 (0.231, 0.319) & \textbf{0.747} (0.721, 0.776) & \textbf{0.344} (0.302, 0.398)\\ 
    Complications of surgical/medical care & acute &0.189 & 0.226 & 0.728 (0.703, 0.752) & 0.464 (0.420, 0.513)  &0.722 (0.698, 0.747) & 0.439 (0.395, 0.487) \\ 
    \textbf{Conduction disorders} & mixed & 0.100 & 0.115 & 0.719 (0.688, 0.750) & 0.252 (0.210, 0.304)  & \textbf{0.854} (0.822, 0.882) & \textbf{0.632} (0.570, 0.692)\\ 
    \textbf{Congestive heart failure; nonhypertensive} & mixed &0.255 &0.295 & 0.760 (0.738, 0.781) & 0.592 (0.553, 0.632)  &\textbf{0.823} (0.805, 0.843) & \textbf{0.679} (0.643, 0.715) \\ 
    \textbf{Coronary atherosclerosis and related} & chronic & 0.311 &0.337 & 0.740 (0.719, 0.763) & 0.603 (0.563, 0.643) & \textbf{0.779} (0.760, 0.799) & \textbf{0.631} (0.593, 0.668)\\ 
    Diabetes mellitus with complications & mixed & 0.114 &0.120 & 0.885 (0.866, 0.902) & 0.534 (0.473, 0.596) &0.883 (0.862, 0.902) & 0.534 (0.473, 0.599)\\ 
    Diabetes mellitus without complication & chronic & 0.172 &0.211 & 0.758 (0.731, 0.781) & 0.430 (0.386, 0.481)  &0.748 (0.724, 0.772) & 0.414 (0.370, 0.463)\\ 
    \textbf{Disorders of lipid metabolism} & chronic & 0.404 &0.406 & 0.689 (0.666, 0.713) & 0.598 (0.562, 0.635)& \textbf{0.707} (0.685, 0.729) & \textbf{0.613} (0.577, 0.649) \\ 
    \textbf{Essential hypertension} & chronic & 0.418 &0.433 & 0.678 (0.655, 0.699) & 0.617 (0.583, 0.650) & \textbf{0.703} (0.682, 0.725) & \textbf{0.634} (0.600, 0.667) \\ 
    Fluid and electrolyte disorders & acute & 0.371 &0.454 & 0.737 (0.716, 0.757) & 0.696 (0.666, 0.727) & 0.733 (0.713, 0.754) & 0.687 (0.657, 0.720) \\ 
    Gastrointestinal hemorrhage & acute & 0.070 &0.071 &0.751 (0.712, 0.785) & 0.194 (0.145, 0.254)& 0.747 (0.708, 0.783) & 0.221 (0.165, 0.287) \\ 
    \textbf{Hypertension with complications} & chronic & 0.215 & 0.222 & 0.736 (0.714, 0.758) & 0.430 (0.391, 0.475)  & \textbf{0.764} (0.742, 0.786) & \textbf{0.465} (0.421, 0.511)\\ %and secondary hypertension
    \textbf{Other liver diseases} & mixed &0.125 &0.169 & 0.716 (0.687, 0.743) & 0.359 (0.313, 0.409)   & \textbf{0.730} (0.704, 0.759) & \textbf{0.398} (0.353, 0.450)\\ 
    Other lower respiratory disease & acute & 0.095 &0.126 & 0.610 (0.574, 0.645) & 0.194 (0.164, 0.238)  &  0.599 (0.564, 0.637) & 0.176 (0.150, 0.210) \\ 
    Other upper respiratory disease & acute & 0.048 &0.054 & 0.746 (0.692, 0.796) & 0.254 (0.185, 0.340)  & 0.753 (0.705, 0.798) & 0.204 (0.148, 0.286)    \\ 
    \textbf{Pleurisy; pneumothorax; pulmonary collapse} & acute & 0.067 &0.095 & 0.627 (0.590, 0.661) & 0.152 ( 0.188, 0.121) & \textbf{0.752} (0.720, 0.783) & \textbf{0.212} (0.174, 0.262) \\ 
    \textbf{Pneumonia} & acute & 0.127 & 0.185 &0.765 (0.738, 0.789) & 0.416 (0.374, 0.470) & \textbf{0.790} (0.766, 0.813) & \textbf{0.453} (0.407, 0.501)\\ 
    Respiratory failure; insufficiency; arrest (adult) & acute & 0.160& 0.282 & 0.845 (0.827, 0.863) & 0.678 (0.637, 0.721) & 0.836 (0.817, 0.854) & 0.653 (0.612, 0.693) \\ 
    Septicemia (except in labor) & acute & 0.158 &0.227 & 0.813 (0.794, 0.834) & 0.572 (0.528, 0.621)&0.809 (0.790, 0.830) & 0.564 (0.522, 0.613)\\ 
    Shock & acute & 0.123 &0.174 & 0.865 (0.844, 0.884) & 0.617 (0.565, 0.666)  & 0.864 (0.843, 0.883) & 0.604 (0.552, 0.654) \\

  \midrule
%   Average acute &  & 0.756 & 0.421 & \textbf{0.767} & \textbf{0.423}\\
%   Average mixed & & 0.734 & 0.447 & \textbf{0.800} & \textbf{0.564} \\
%   Average chronic & & 0.720 & 0.487 & \textbf{0.750} & \textbf{0.517 }\\
%   \midrule
% Average & all & 0.742 $\pm0.027$ & 0.445 $\pm0.046$ & \textbf{0.770} $\pm0.026$  & \textbf{0.477} $\pm0.046$ \\
  Average & all & - & - & 0.746 (0.720, 0.772)& 0.453 (0.409, 0.502)  & \textbf{0.770} (0.745, 0.795)  & \textbf{0.481} (0.436, 0.531) \\
  \bottomrule 
    \end{tabular}}
    \label{tab:phenotype_wise}
\end{table*}

\begin{table*}[ht!]
    \centering
    \caption{\small\textbf{Performance of \texttt{MedFuse} across different age groups for in-hospital-mortality on the $(\mathbf{EHR}+\mathbf{CXR})_{\mathbf{PAIRED}})$ test set, compared to the uni-modal stacked LSTM with $\mathbf{EHR}_{\mathbf{PAIRED}}$.} We compare the AUROC and AUPRC for the different age groups. The results in bold indicate improved performance with multi-modal data.}\vspace{-2mm}
     \resizebox{1.0\textwidth}{!}{\begin{tabular}{l c c c| c c c} \toprule
     & & \multicolumn{ 2}{c}{$\textbf{EHR}_\textbf{PAIRED}$} &  \multicolumn{ 2}{c}{$\textbf{(EHR+CXR)}_\textbf{PAIRED}$} &  \\
     \midrule
    
    \textbf{Age group} & \textbf{Positive fraction} & \textbf{AUROC} & \textbf{AUPRC} & \textbf{AUROC} & \textbf{AUPRC} \\
    \midrule

    18-40 &  0.078 (11/141) & 0.941 (0.859, 0.988) & 0.613 (0.332, 0.883) & 0.917 (0.820, 0.980) & \textbf{0.521} (0.272, 0.841) \\

    40-60 &  0.119 (44/369) & 0.796 (0.719, 0.865) & 0.403 (0.277, 0.541) & \textbf{0.822} (0.751, 0.885) & \textbf{0.499} (0.360, 0.629) \\
    
    60-80 & 0.159 (98/616) & 0.846 (0.805, 0.885) & 0.576 (0.478, 0.670) & \textbf{0.868} (0.830, 0.900) & \textbf{0.583} (0.484, 0.678) \\
    
    $>$ 80 & 0.227 (56/247) & 0.789 (0.722, 0.850) & 0.549 (0.428, 0.677)  & \textbf{0.841} (0.784, 0.891) & \textbf{0.616} (0.491, 0.731) \\

%   \midrule
%   Average & all & 0.742 & 0.445 & \textbf{0.770}  & \textbf{0.477} \\
  \bottomrule
    \end{tabular}}
    \label{tab:age_analysis}
\end{table*}

\subsection{Phenotype-wise Analysis}
In Figure~\ref{fig:types_bar}, we show the AUROC (left) and AUPRC (right) results across different categories of phenotype labels: acute, mixed, and chronic conditions. The label types and their prevalence are listed in Table~\ref{tab:phenotype_wise}. We note that our approach mostly improves the performance in terms of AUROC and AUROC for mixed and chronic conditions, which are generally hard to predict through uni-modal clinical time-series data~\citep{benchhmark}. In particular, across mixed conditions, the AUROC increases from 0.749 to 0.800, and the AUPRC increases from 0.458 to 0.565. For chronic conditions, the AUROC increases from 0.717 to 0.745 and the AUPRC increases from 0.487 to 0.512. We observe relatively smaller improvements for acute conditions, where the AUROC increases from 0.761 to 0.772 and the AUPRC increases from 0.432 to 0.433. In Table~\ref{tab:phenotype_wise}, we report the performance across all 25 labels for the paired test set using uni-modal and multi-modal data. We observe an improvement across a number of thorax-related phenotypes, such as pneumonia and pleurisy, which are usually clinically assessed using chest imaging~\citep{long2017emergency}. This further highlights the importance of using the chest X-ray images as auxiliary information along with the clinical time-series data.

\vspace{-2mm}
\subsection{In-hospital Mortality Age-wise Analysis}
We evaluate the performance of our approach across different age groups, as shown in  Table~\ref{tab:age_analysis}, and compare it to the uni-modal stacked LSTM. We observe that the AUROC and AUPRC improve across age groups 40-60, 60-80, and $>$80 years, while the AUROC decreases for the 18-40 years. The latter result needs further investigation with a larger dataset, since the test sets only contain 11 positive samples for the youngest age group. Additionally, there are variations in the relative improvements. For example, the AUPRC increases by 24\% for the 40-60 years group, compared to 1.3\% in the 60-80 years group. %This highlights that multi-modal learning can have more significant gains for certain age groups. 

% \vspace{-2mm}
% \subsection{Instance-level Analysis}

\section{Discussion}
% Summary of main findings
In this paper, we present a multi-modal fusion approach, named \texttt{MedFuse}, and new benchmark results for integrating partially paired clinical time-series data and chest X-ray images. We evaluate it for two popular benchmark tasks, namely in-hospital mortality prediction and phenotype classification, using publicly available datasets MIMIC-IV and MIMIC-CXR.

Our study has several strengths. First, our approach is simple and easy to implement. The results show that the proposed approach performs better than the uni-modal LSTM baseline, as it considers chest X-ray images, when available, as an additional source of information. In addition, the approach outperforms several baselines, and the phenotype-wise and age-wise analysis provide some insight as to where it improves performance. We conclude that the proposed method is overall a better choice than the baseline methods because (i) the LSTM-based fusion module can inherently deal with missingness (i.e., partially paired data), and (ii) the combination of the architecture and the training procedure provides performance gains. Otherwise, the size of the partially paired training set does not seem to be correlated with the performance improvements, as illustrated with the validation set results in Appendix~\ref{unimodalpercentagetraining}. The results overall highlight the promise of multi-modal fusion in improving the performance of clinical prediction models. Multi-modal learning is also generally more closely aligned with the decision-making process of clinicians, who consider multiple sources of information when assessing a patient.

Moreover, in contrast with conventional multi-modal approaches that assume paired input, our proposed method is more flexible since it can process samples with missing chest X-ray images. There is a rising interest in learning cross-modal interactions between modalities during training time and in reconstructing missing modalities~\citep{ngiam2011multimodal, 9534148,missing, sylvain2021cmim,missing}. In contrast with natural multi-modal datasets, assuming a high degree of correlation in such settings is not a trivial task in healthcare especially when the modalities do not necessarily share the same labels, and this is an area of future work. The difficulty stems from the sparse and asynchronous nature of medical data, i.e. it would be difficult to use a biopsy report for skin tissue to reconstruct common thorax diseases features~\citep{hayat2021dynamic}. Additionally, some of the existing work for learning cross-modal interactions assumes the presence of all modalities during training~\citep{sylvain2021cmim}.

Another strength is that the approach can be easily scaled to more than two modalities with no amendments to the fusion loss function, compared to existing work where the complexity of the computation increases with the number of modalities~\citep{hayat2021dynamic}. However, this requires evaluation and is an area of future work. We also do not assume any correlation among the input modalities, in terms of information content or assigned labels.

Furthermore, we formalize and introduce new benchmark results for two popular tasks that are typically evaluated in the context of clinical time-series data only~\citep{benchhmark}. By gaining access to the MIMIC-IV and MIMIC-CXR datasets~\citep{mimic4,mimiccxrjpg}, researchers can utilize our open-access data pre-processing pipeline and introduce new results for direct comparison.

% From a clinical perspective... 

% Our results show that the inclusion of the images provides performance gains for both tasks at inference time, especially for multi-modal samples. The inclusion of the additional modality is relatively more significant for younger patients (40-60 years old), highlighting the potential role of partially paired multi-modal learning in creating fairer predictive models.

\paragraph{Limitations.} The study also has its own limitations. To begin with, we focus on tasks pertaining to the integration of clinical time-series data and chest X-ray images from a single data source, and we evaluate our work on two benchmark tasks due to limited resources. The original work by~\cite{benchhmark} includes two other tasks, decompensation prediction and length of stay prediction, which we would like to evaluate our method on in the future. The in-hospital mortality task should also be investigated in the setting where chest X-rays collected beyond the first 48 hours of the ICU stay are excluded. We also do not run any experiments on settings where the clinical time-series data may be missing, but the chest X-ray image is available. In future work, this requires the definition of additional benchmark tasks where the chest X-ray image is the primary modality.  Since we currently evaluate our method with two input modalities only, another interesting next step would be to use more than two to further evaluate the robustness of the model, considering its scalability. In its current formulation, the model also lacks interpretability, since we mainly focus on fusion within the scope of this paper. We later plan to explore incorporating attention layers~\citep{vaswani2017attention} at the input level of the feature encoders to evaluate the importance of features within each modality, and within the fusion module to evaluate the overall informativeness of each modality. On a related note, our work can benefit from performing instance-level analysis. However, this requires clinical expertise that bridges between chest X-ray image and clinical time-series analysis, which we are currently missing. To realize the full potential of multi-modal learning, there is more work to be done to understand the clinical underpinnings of multi-modal fusion. Overall, the study highlights an extremely worthwhile direction to further leverage the value of multi-modal learning in healthcare, especially as the diversity and quantity of medical data continues to increase. 

%Since this requires additional data pre-processing overhead, we leave it for future work.
% - cross validation (currently one split of data)
% -instance level analysis (requires clinical expertise with both modalities)
% Limitation 1: data is restricted 

% \section{Conclusion} To summarize, we propose a simple yet novel and promising multi-modal fusion approach, \texttt{MedFuse}. Our extensive experiments and analysis with a publicly available dataset for two important clinical prediction tasks show that our approach performs favorably against several baselines. 

\paragraph{Acknowledgements.} This work is supported in part by the NYUAD Center for Artificial Intelligence and Robotics, funded by Tamkeen under the NYUAD Research Institute Award CG010. We would also like to thank the High Performance Computing (HPC) team at NYUAD for their support.

\bibliography{main}

\newpage
 \appendix
 \section{}
 \label{appendix-a}
 \setcounter{table}{0}
\renewcommand{\thetable}{A\arabic{table}}

 \setcounter{figure}{0}
\renewcommand{\thefigure}{A\arabic{figure}}
 
\subsection{Image Augmentations}
\label{image-aug}
For the chest X-ray images, we apply a series of transformations during pre-training and fine-tuning across all experiments and tasks. Specifically, we resize each image to $256 \times 256$ pixels, randomly apply a horizontal flip, and apply a set of random affine transformations, such as rotation, scaling, shearing, and translation. We then apply a random crop to obtain an image of size $224 \times 224$  pixels. During validation and testing, we perform image resizing to $256 \times 256$ and apply a center crop to $224 \times 224$ pixels.

\subsection{Hyperparameter Search Results}
\label{hyperparameter} 
 The results of hyperparameter tuning are shown in Table~\ref{tab:learning_rates}. We summarize the learning rates that achieved the best performance for each model.

\begin{table}[h!]
    \centering
    \vspace{-5mm}
    \caption{\small{\textbf{Learning rates that achieved the best results during hyperparameter search.} We conducted 10 runs for each model with randomly sampled learning rates between $10^{-5}$ and $10^{-3}$. For MMTM and DAFT, we additionally selected the version that achieved the best validation set AUROC.}} \vspace{-1mm}
     \resizebox{0.85\textwidth}{!}{\begin{tabular}{l c | c } \toprule
     Task & \textbf{Phenotyping} &  \textbf{In-hospital mortality} \\
     \midrule
    \textbf{Method}  & \multicolumn{2}{c}{\textbf{Learning rate}} \\
    \midrule
       LSTM trained with $\mathbf{EHR}_{\mathbf{PAIRED}}$ & $8.866 \times 10^{-5}$ & $1.000\times 10^{-4}$\\
       LSTM trained with $\mathbf{EHR}_{\mathbf{PARTIAL}}$ & $5.399\times 10^{-4}$ & $5.399\times 10^{-4}$\\
       Early trained with ($\mathbf{EHR+CXR})_{\mathbf{PARTIAL}}$ & $9.084\times 10^{-5}$ &  $ 3.095 \times 10^{-4}$ \\
       Early trained with ($\mathbf{EHR+CXR})_{\mathbf{PAIRED}}$ & $3.833\times 10^{-5}$ & $9.515\times 10^{-5}$ \\
       Joint trained with ($\mathbf{EHR+CXR})_{\mathbf{PARTIAL}}$ & $3.831\times 10^{-5}$ & $7.565 \times 10^{-4}$ \\
       Joint trained with ($\mathbf{EHR+CXR})_{\mathbf{PAIRED}}$ & $5.652\times 10^{-5}$ & $4.032\times 10^{-5}$ \\
       MMTM$^*$ & $5.326\times 10^{-5}$ & $4.355\times 10^{-5}$\\
       DAFT$^{**}$ & $6.493\times 10^{-5}$ & $6.493\times 10^{-5}$ \\
       Unified & $2.042\times 10^{-4}$ & $2.606\times 10^{-4}$ \\
    %   \midrule
    \texttt{MedFuse} (Randomly initialized encoders) & $4.741\times 10^{-5}$ & $9.382\times 10^{-5}$ \\
       \texttt{MedFuse} (Pre-trained encoders) & $7.347\times 10^{-5}$ & $1.452\times 10^{-5}$ \\
       
       \bottomrule
       \multicolumn{3}{p{15cm}}{\footnotesize{$^{*}$We trained two versions of MMTM for each task, where the MMTM module is placed after the third or fourth ResNet layer. Placing it after the fourth layer achieved the best performance for both tasks.}} \\
       \multicolumn{3}{p{15cm}}{\footnotesize{$^{**}$We trained two versions of DAFT for each task, where we transform the LSTM representation either after the third or fourth ResNet layer. Placing it after the third layer achieved the best performance for phenotype classification, whereas placing it after the fourth layer achieved the best performance for in-hospital mortality.}} \\
    \end{tabular}}
    \label{tab:learning_rates}
\end{table}

\subsection{Percentage of Uni-modal Samples within the Training Set} %\vspace{-2mm}
\begin{figure}[h!]
    \centering
    \vspace{-1mm}
    \includegraphics[width=.5\textwidth]{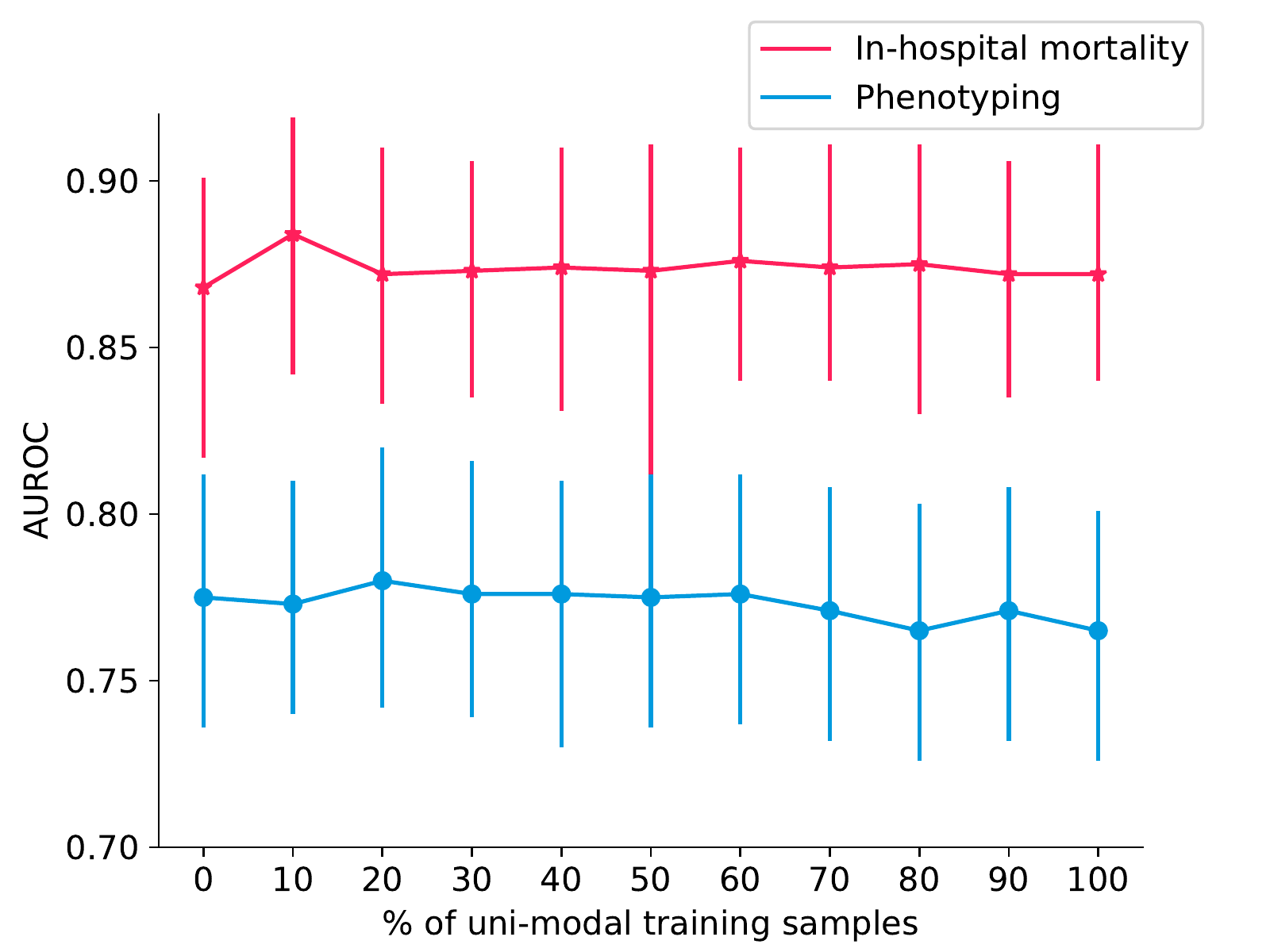}
    \vspace{-3mm}
    \caption{\small\textbf{Performance on the validation set when varying the sampling percentage for uni-modal training samples.} The plot shows the AUROC on the validation set for different percentages of randomly selected uni-modal training samples. }%We trained the model with the selected uni-modal samples and all of the available paired samples during a single training epoch of \texttt{MedFuse}. 
    \label{fig:data_ratio}
    \vspace{-4mm}
\end{figure}

\label{unimodalpercentagetraining}
We also run experiments where we vary the percentage of uni-modal samples during fine-tuning. The best AUROC results for both tasks on the validation set are shown in Figure~\ref{fig:data_ratio}. For in-hospital mortality (shown in red), we notice that a relatively smaller portion of uni-modal samples (10\%) achieves the best performance. For patient phenotyping (shown in blue), we observe a similar trend where the best AUROC is achieved with only 20\% of uni-modal samples. We fix the sampling percentage that achieves the best validation AUROC across all experiments, unless noted otherwise. Hence, this highlights that the best performance gains of \texttt{MedFuse} are achieved even with a small percentage of uni-modal samples.

\subsection{Percentage of Uni-modal Samples within the Paired Test Set}
\label{unimodalpercPAIRED}
\begin{figure}[h!]
    \centering
    \vspace{-1mm}
    \includegraphics[width=.6\textwidth]{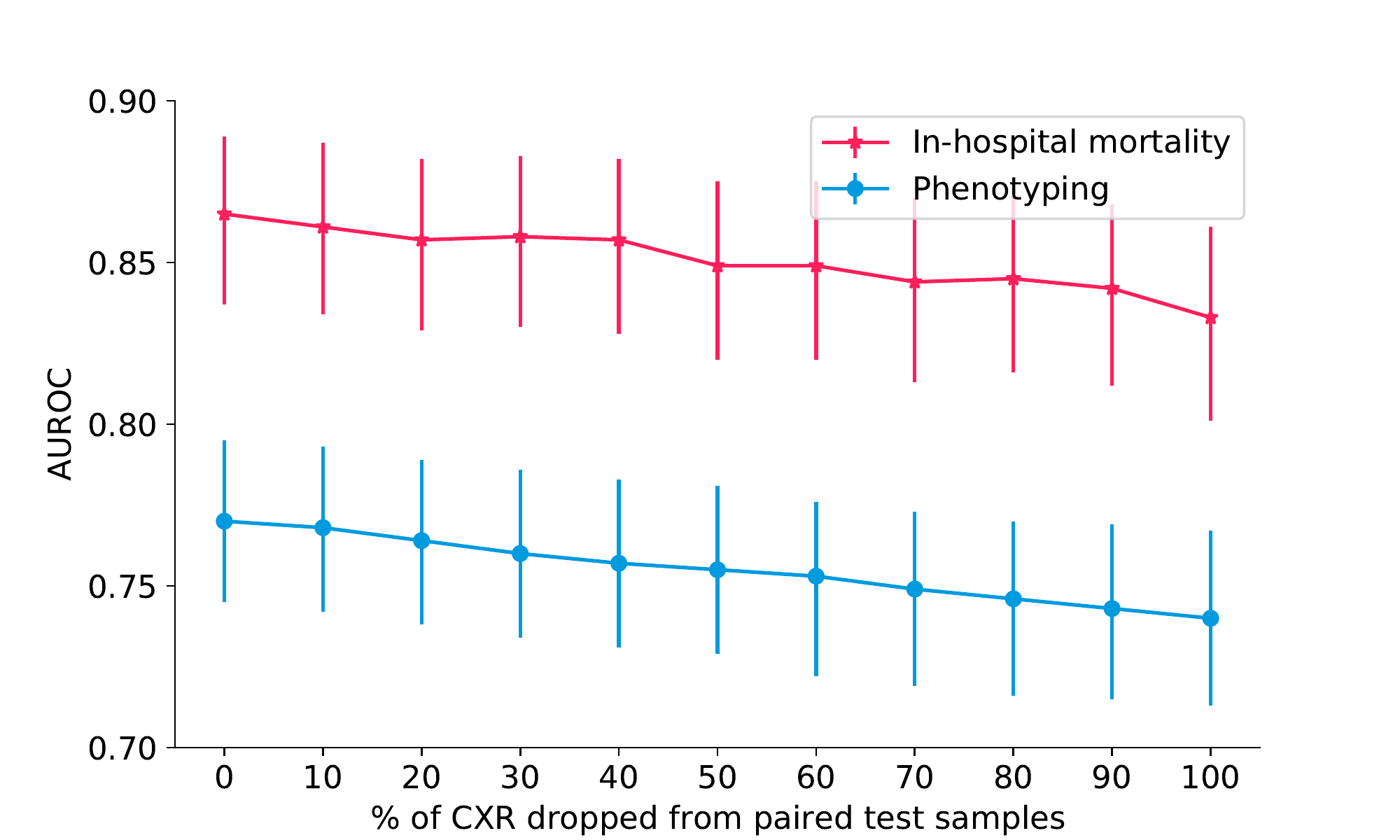}
    \vspace{-3mm}
    \caption{\small\textbf{Performance on the test set with randomly dropped CXR modality in the paired test set.} The plot shows the AUROC on the paired test set for different percentages of randomly dropped CXR modality from paired test samples.}
         \label{fig:data_ratio_paired}
    \vspace{-4mm}
\end{figure}

We also performed an ablation study where we randomly dropped the chest X-ray modality for a percentage of samples in the paired test set. The results are shown in Figure~\ref{fig:data_ratio_paired}. We observe that the as the percentage of dropping increases, the AUROC decreases for both tasks.

\newpage
\subsection{Missing Modality with Early and Joint Fusion} %\vspace{-2mm}
\label{missingtoken}
We also ran initial experiments to compare the learnable vector with imputing zeros for a missing chest X-ray modality. The results are shown in Table~\ref{tab:missing-modality}. We note that the results are comparable with no obvious differences.

\begin{table*}[h!]
    \centering
    \caption{\textbf{Missing modality with early and joint fusion.} We report the AUROC and AUPRC results on the entire test set ($\mathbf{EHR}_{\mathbf{PARTIAL}}$), including samples with missing chest X-ray images (substituted with a zeros or a learnable vector) All methods below were pre-trained using the ($\mathbf{EHR}_{\mathbf{PARTIAL}}$) training set and a fixed learning rate of 0.0001.} \vspace{-1mm}
     \resizebox{0.75\textwidth}{!}{\begin{tabular}{l c c c| c c c} \toprule
     \multicolumn{ 2}{c}{\textbf{Task}} &  \multicolumn{ 2}{c}{\textbf{Phenotyping}} &  \multicolumn{ 2}{c}{\textbf{In-hospital mortality}} \\
     \midrule
    
    \textbf{Method} & \textbf{Missing Vector}  & \textbf{AUROC} & \textbf{AUPRC} & \textbf{AUROC} & \textbf{AUPRC} \\
    \midrule
       
      Joint & Zeros & 0.756  & 0.406  & 0.843  & 0.466  \\
      Joint & Learnable & 0.752& 0.402 & 0.853  & 0.486 \\
      Early & Zeros & 0.743  & 0.392  & 0.842  & 0.481 \\
      Early & Learnable & 0.742 & 0.388  & 0.851  & 0.489 \\
    %   \midrule
    %   \texttt{MedFuse} & Zeros & \textbf{0.765} $\pm0.012$ & \textbf{0.421} $\pm0.022$ & \textbf{0.869} $\pm0.014$ & \textbf{0.552} $\pm0.036$ \\
       
      \bottomrule
    \end{tabular}}
    \label{tab:missing-modality}
\end{table*}

\subsection{Percentage of Uni-modal Samples within the Partially Paired Test Set}
\label{unimodalpercPARTIAL}
\begin{figure}[h!]
    \centering
    % \vspace{-1mm}
    \includegraphics[width=.6\textwidth]{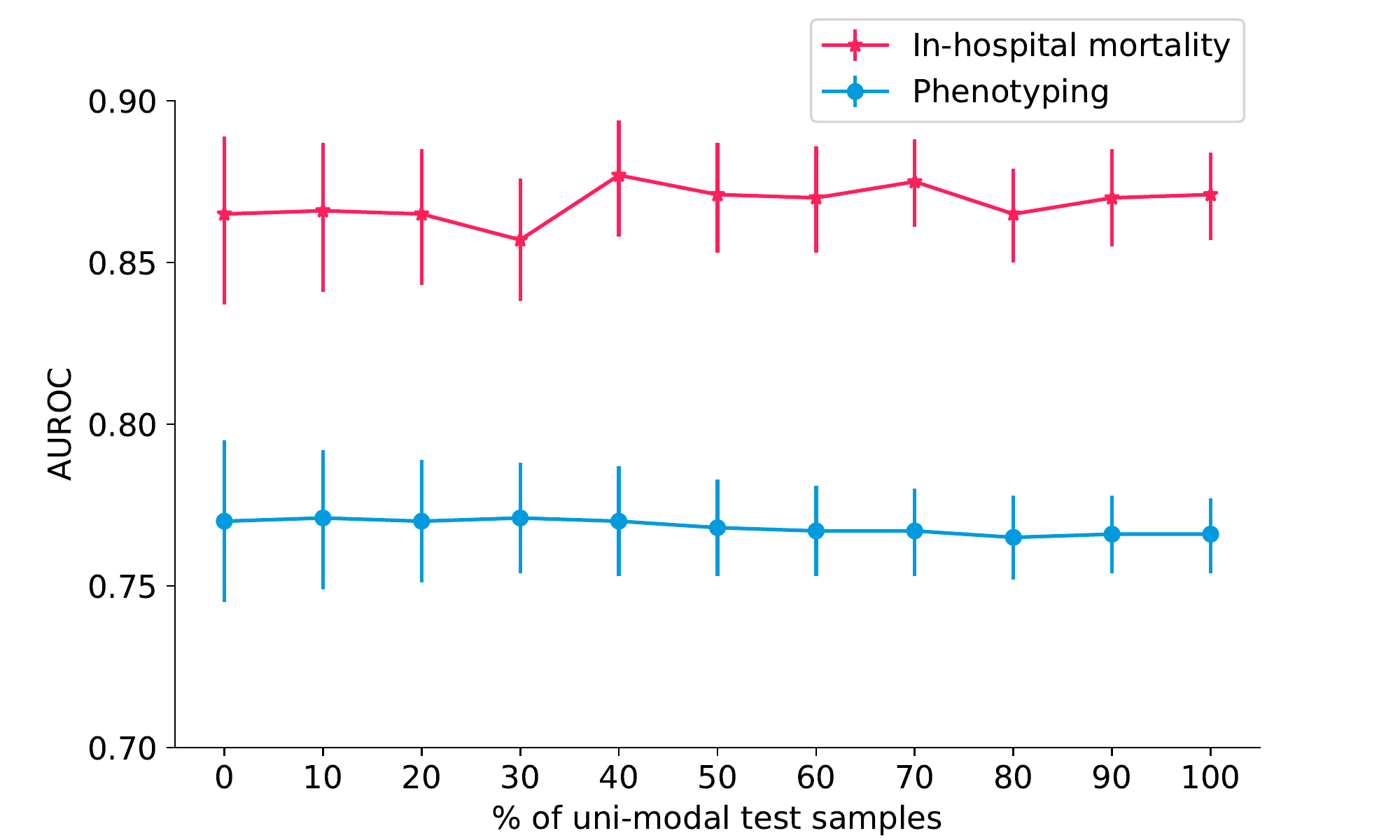}
         \caption{\small\textbf{Performance on the test set when varying the percentage of uni-modal samples.} The plot shows the AUROC on the partial test set for different percentages of randomly selected uni-modal samples.}
         \label{fig:data_ratio_pairtial_testset}
    \vspace{-4mm}
\end{figure}

We performed another ablation study where we varied the number of uni-modal samples in the partially paired test set. The results are shown in Figure~\ref{fig:data_ratio_pairtial_testset}. Hence, including 0\% of uni-modal test samples is equivalent to the fully paired test set. We observe an increase in the AUROC in the in-hospital mortality task, as the percentage of uni-modal samples increase, but a more a consistent AUROC in the phenotyping task. We do however observe that the widths of the confidence intervals decrease as the percentage of uni-modal samples increases across both tasks. 

% \usepackage{subcaption}
% \begin{figure}[]
%      \centering
%      \begin{subfigure}{.5\textwidth}
%          \centering
%          \includegraphics[width=0.8\linewidth]{figures/aurocs_paired_ratio.pdf}
%          \caption{\small\textbf{Performance on the test set with randomly dropped percentage of CXR modality from paired test set.} The plot shows the AUROC on the paired test set for different percentages of randomly dropped CXR modality from paired test samples.}
%          \label{fig:data_ratio_paired}
%      \end{subfigure}
%      \hfill
%      \begin{subfigure}{.5\textwidth}
%          \centering
%          \includegraphics[width=0.8\linewidth]{figures/aurocs_nonpaired_ratio.pdf}
%          \caption{\small\textbf{Performance on the test set when varying the sampling percentage for uni-modal samples.} The plot shows the AUROC on the partial test set for different percentages of randomly selected uni-modal test samples. }
%          \label{fig:data_ratio_pairtial_testset}
%      \end{subfigure}
% \end{figure}

\subsection{Ensemble of uni-modal and multi-modal models} %\vspace{-2mm}
\label{ensemble}
We ran another experiment to compare the performance of \texttt{MedFuse} to that of an ensemble of two models: an EHR only model that computes predictions for partial input (i.e., not associated with a chest X-ray) using LSTM, and a paired model that computes predictions for paired input using \texttt{MedFuse}. The results are shown in Table~\ref{tab:medfuse_unimodal}. We observe that the ensemble slightly outperforms \texttt{MedFuse} for phenotyping only. This implies that an ensemble of strong models may be better suited for some tasks, such as phenotyping, which however requires the training of two models.

\begin{table*}[h!]
    \centering
    \caption{\textbf{MedFuse compared to an ensemble evaluation.} We report the AUROC and AUPRC results on the partially paired test set.} \vspace{-1mm}
     \resizebox{0.75\textwidth}{!}{\begin{tabular}{l c c| c c c} \toprule
    \textbf{Task}&  \multicolumn{ 2}{c}{\textbf{Phenotyping}} &  \multicolumn{ 2}{c}{\textbf{In-hospital mortality}} \\
     \midrule
    
    \textbf{Method} & \textbf{AUROC} & \textbf{AUPRC} & \textbf{AUROC} & \textbf{AUPRC} \\
    \midrule
       
      Ensemble & 0.770 (0.759, 0.782) & 0.431 (0.410, 0.454)  & 0.870 (0.857, 0.884)  & 0.547 (0.509, 0.589) \\
      \texttt{MedFuse} & 0.768 (0.756, 0.779) & 0.429 (0.408, 0.452) & 0.874 (0.860, 0.888)& 0.567 (0.529, 0.607) \\
      \bottomrule
    \end{tabular}}
    \label{tab:medfuse_unimodal}
\end{table*}

% We ran evaluation for uni-modal and multi-modal on paired dataset. The results are shown in Table~\ref{tab:medfuse_unimodal}. We note that the results are comparable with no obvious differences.  

% \begin{table*}[h!]
%     \centering
%     \caption{\textbf{MedFuse uni-modal and multi-modal evaluation.} We report the AUROC and AUPRC results on the paired test set ($\mathbf{EHR}_{\mathbf{PAIRED}}$). The MedFuse was traind using partially paired training set.} \vspace{-1mm}
%      \resizebox{0.75\textwidth}{!}{\begin{tabular}{l c c| c c c} \toprule
%     \textbf{Task}&  \multicolumn{ 2}{c}{\textbf{Phenotyping}} &  \multicolumn{ 2}{c}{\textbf{In-hospital mortality}} \\
%      \midrule
    
%     \textbf{Method} & \textbf{AUROC} & \textbf{AUPRC} & \textbf{AUROC} & \textbf{AUPRC} \\
%     \midrule
       
%       CXR & 0.537 (0.526, 0.548)  & 0.214 (0.204, 0.226)  & 0.546 (0.527, 0.564  &  0.161 (0.143, 0.182)   \\
%       EHR & 0.740 (0.713, 0.767) &  0.441 (0.398, 0.489) &  0.833 (0.802, 0.861) & 0.514 (0.443, 0.584) \\
%       EHR+CXR & 0.770 (0.745, 0.795)  & 0.481 (0.436, 0.531)  & 0.865 (0.837, 0.889)  & 0.594 (0.526, 0.655) \\
%     %   \midrule
%     %   \texttt{MedFuse} & Zeros & \textbf{0.765} $\pm0.012$ & \textbf{0.421} $\pm0.022$ & \textbf{0.869} $\pm0.014$ & \textbf{0.552} $\pm0.036$ \\
       
%       \bottomrule
%     \end{tabular}}
%     \label{tab:medfuse_unimodal}
% \end{table*}

\end{document}